\documentclass[proof]{WileyASNA-v1}

\articletype{Article Type}%

\received{14 July 2025}
\revised{12 September 2025}
\accepted{22 October 2025}

\begin{document}

\title{Unveiling the Cosmos: XMM-Newton's Scientific Strategy 
}

\author[1]{Norbert Schartel}

\author[1]{Maria Santos-Lleo}

\authormark{Schartel \textsc{et al}}

\address[1]{\orgdiv{ESAC}, \orgname{ESA}, \orgaddress{Camino Bajo del Castillo s/n, 28692 Villanueva de la Ca{\~a}da\state{Madrid}, \country{Spain}}}

\corres{*Norbert Schartel \email{Norbert.Schartel@gmail.com}}

\abstract{In December 2024, the European Space Agency's (ESA)  XMM-Newton X-ray Observatory celebrated  the 25th anniversary of its launch. The annual number of peer-reviewed articles utilising XMM-Newton data has exhibited a consistent upward trajectory over the past two and a half decades, attaining more than 400 in 2022. The annual call for observing time proposals continues to experience  a  high level of oversubscription, typically ranging from a factor of 6 to 7.
In order to  enhance the scientific discovery space, XMM-Newton, 
primarily through the 
Project Scientist and Science Operations Centre, has pursued a strategy of expansion, which can be grouped into three phases:  Large Projects with long observing time (2006-2009), Joint Observations (2011-2016), and Targets of Opportunity (2016-2024), respectively. 
A salient feature of XMM-Newton's time allocation is the systematic removal of biases from the second call onwards, a strategy that has enabled the attainment of comparable gender success rates and high acceptance rates for young scientists over 25 years, a feat only recently accomplished by similar missions through the introduction of double-anonymous review. XMM-Newton research is conducted by an active community of 4,300 scientists, of which approximately 570 are leading (1st author). 
The foundation of this community and its research is predicated on XMM-Newton data, with the project's policy of user support and calibration being fundamental constituents, as well as the project's active engagement and communication with its members.

}

\keywords{XMM-Newton}

%\keywords{ \LaTeXe; \emph{Wiley NJD}}

\ctitle{Unveiling the Cosmos: XMM-Newton's Scientific Strategy, \cjournal{Astronomical Notes}, \cvol{2025}.}

\maketitle

\footnotetext{\textbf{Abbreviations:} ESA, European Space Agency}

\section{Introduction}

The X-ray Multi-Mirror Mission, XMM-Newton \citep{2001A&A...365L...1J, 2022hxga.book..114S} 
represents the second cornerstone of the European Space Agency's (ESA) Horizon 2000 Science Programme, providing an observatory-class X-ray facility.
XMM-Newton was launched on 10 December 1999, aboard an Ariane 5 rocket, and is expected to extend well beyond 2032 at the current rate of fuel consumption. 

The observatory is capable of simultaneous non-dispersive spectroscopic imaging and timing, utilising the European Photon Imaging Camera (EPIC) \citep{2001A&A...365L..27T,2001A&A...365L..18S}, consisting of one pn-camera and two MOS-cameras.
Moreover, the observatory provides simultaneous medium-resolution dispersive X-ray spectroscopy (Reflection Grating Spectrometer; RGS \citep{2001A&A...365L...7D}) and optical/UV imaging, spectroscopy and timing from a co-aligned telescope (Optical Monitor; OM \citep{2001A&A...365L..36M}).

The design of the XMM-Newton cameras was conceived with the objective of optimising the effective area at 6.4 keV for the purpose of spectroscopic studies. Collectively, the three EPIC cameras provide an effective area throughout the energy range from 300 eV to 12 keV, reaching up to 2500 cm{$^2$} at 1.5 keV and approximately 1800 cm{$^2$} at 5 keV. 
The 30 arcmin diameter field of view of the EPIC cameras permits the study of extended sources, e.g. supernova remnants, galaxies,  or clusters of galaxies.
Each of the two RGS instruments has an effective area of up to 60 cm{$^2$}  at 15 \AA. This range extends from approximately 0.3 -- 2.2 keV.  The design of the RGS was conceived with the intention of covering the most intense emission lines at soft X-ray energies. 

NASA's Chandra X-ray Observatory \citep{2002PASP..114....1W}{Wilkes2020}, launched in 1999,  exhibits superior spatial resolution of its cameras and high-spectral resolution of its gratings, although with a smaller effective area. The two observatories are highly complementary in terms of their respective capabilities, offering a comprehensive range of scientific insights. 

Large space-based observatories are, in principle, second only to large particle facilities in terms of financial investment. Extending their lifetimes, if scientifically justified, is therefore an ethical mandate. Within the ESA operating scheme, the running costs are minimal compared to the initial investment, on the order of one or two percent.
XMM-Newton is operated by ESA's  Science and Mission Operations Centres (SOC and MOC), with the support of the Survey Science Centre Consortium (SSC) and the NASA XMM-Newton Guest Observer Facility (GOF). 
ESA policy is to authorise operations of its missions over a period of three years. Continued
operation for longer periods of time involves a so-called "mission extension" process, through
which, on a triennial basis, a proposal for the prospective operation of XMM-Newton is submitted
to the ESA Science Programme Committee (SPC).

The SPC reaches a decision based on a technical evaluation of the satellite and its ground segment, as well as a scientific rationale presented in the form of a proposal. 
This scientific proposal must clearly delineate the research topics  to be addressed by future XMM-Newton observations.  
In particular, it should identify any novel opportunities that  were not previously possible. While the process of extension is primarily administrative and budgetary in nature, it naturally gives rise to considerations of the mission's scientific benefits and potential  improvements to mission operations and observing strategy. This  ensures XMM-Newton's  continued scientific impact and contribution to the field. The following article provides an overview of how this science strategy has evolved over the past two and a half decades.

The article is structured as follows: First, an explanation is provided of how  scientific publications are monitored and assessed by analysing the temporal evolution of the 
published papers, their authors, and the observations analysed therein.
Second, an exploration is made of the typical characteristics of research groups using XMM-Newton data (the `user model'), with the goal of better accommodating their needs. Third, we discuss  how feedback from an engaged community has been incorporated to improve support. The subsequent section shows how we have ensured a sustained expansion of the discovery space as the mission proceeded. Finally,  we expound on the execution and refinement of the diverse tasks of an observatory-class  mission, with the objective of optimising the scientific output .

\section{Scientific Output and User Community}\label{Out}

To characterise the scientific output of XMM-Newton, two key metrics are monitored: (1) the number of peer-reviewed publications and (2) the oversubscription factor of the annual call for observing proposals, which is defined as the ratio between the time requested by the proposals received and the time expected to be available for observing. 

\begin{figure}[t]
\centerline{\includegraphics[angle=0,scale=0.3]{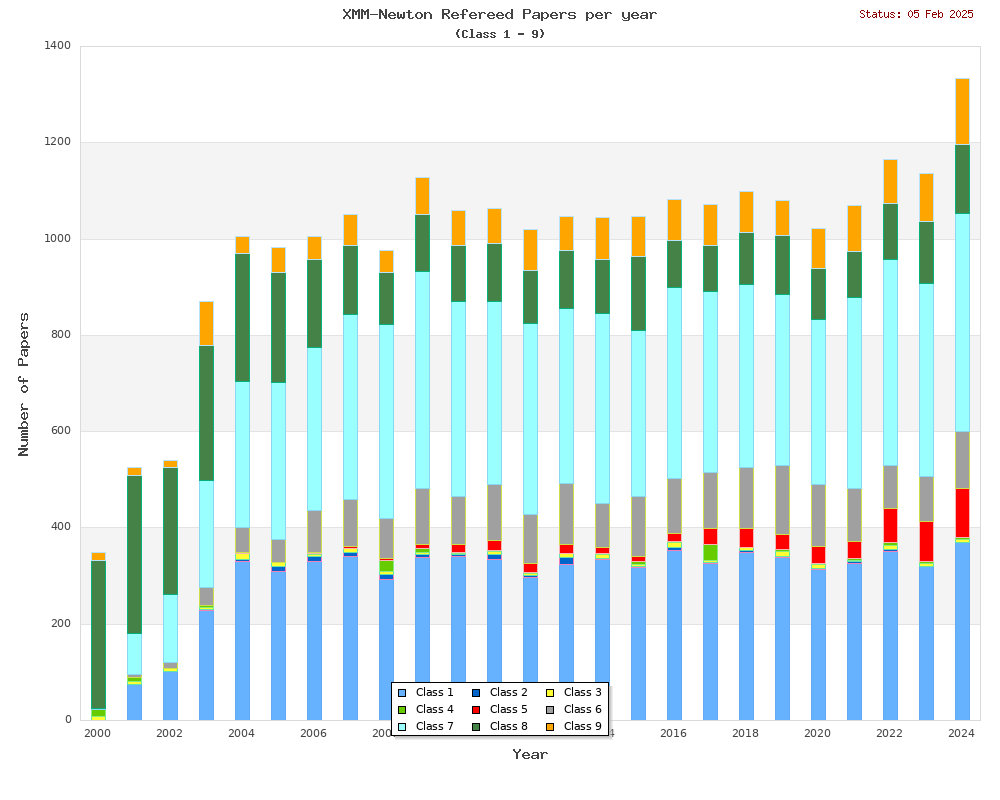}}
\caption{
The number of refereed articles related to XMM-Newton published per year and class.   (See text for details of the classification.) 
\label{fig1}}
\end{figure}

A search of the NASA Astrophysics Data System (ADS) (body text) for  "XMM" yields the refereed articles, which are then reviewed manually and categorised as follows:
\begin{itemize}
    \item Class 1 comprises articles that employ data obtained with XMM-Newton.
\item Class 2 comprises catalogues based on data obtained with XMM-Newton.
\item Class 3 comprises articles that make numerical predictions for XMM-Newton observations.
\item Class 4 comprises articles that describe XMM-Newton, its instruments, and its scientific impact.
\item Class 5 comprises articles that employ products or catalogues provided by the XMM-Newton SOC or SSC.
\item Class 6 comprises articles that utilise  XMM-Newton results published in other articles.
\item Class 7 comprises articles that cite other articles based on XMM-Newton observations.
\item Class 8 comprises articles that make reference to XMM-Newton, but do not cite a specific article.
\item Class 9 comprises articles that either include the term "XMM" in astrophysical source nomenclature or references, or mention it in a context that is not strictly scientific, such as a funding statement.
\end{itemize}
Figure 1 illustrates the annual number of refereed articles involving XMM-Newton, classified according to these  criteria.
Details on the search, manual review, and categorization can be found in \cite{Ness2}.

\begin{figure}[t]
\centerline{\includegraphics[angle=0,scale=0.3]{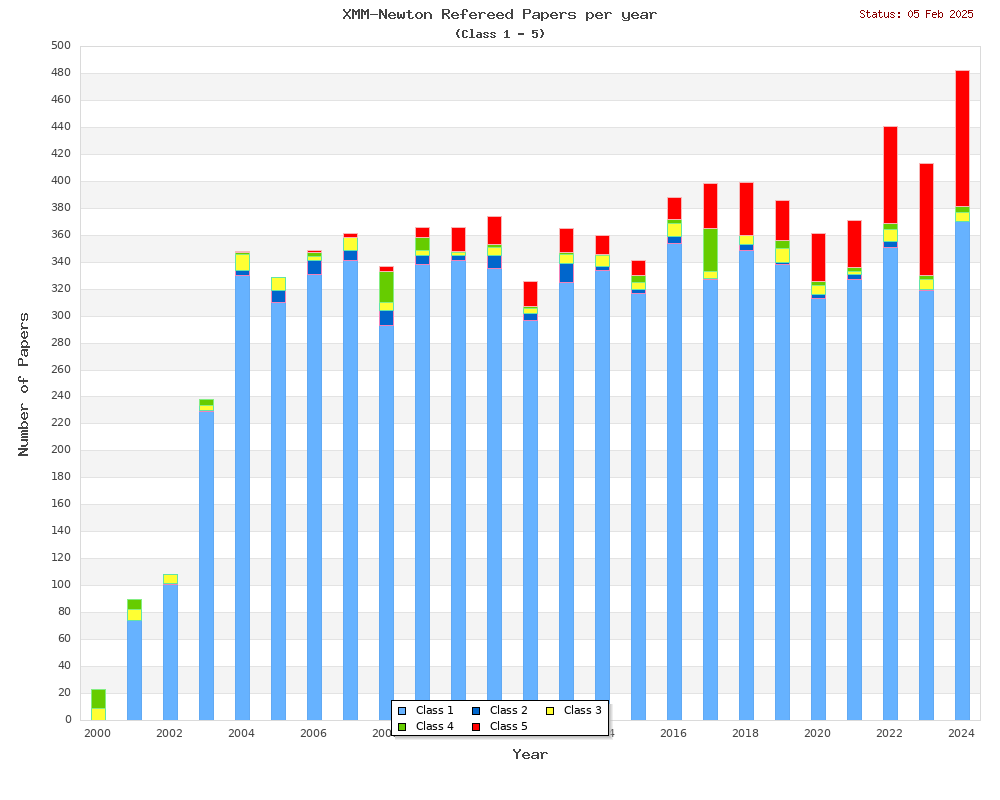}}
\caption{Same as \ref{fig1}, but now only for class 1 to 5.
\label{fig2}}
\end{figure}

\begin{figure}[t]
\centerline{\includegraphics[angle=0,scale=0.25]{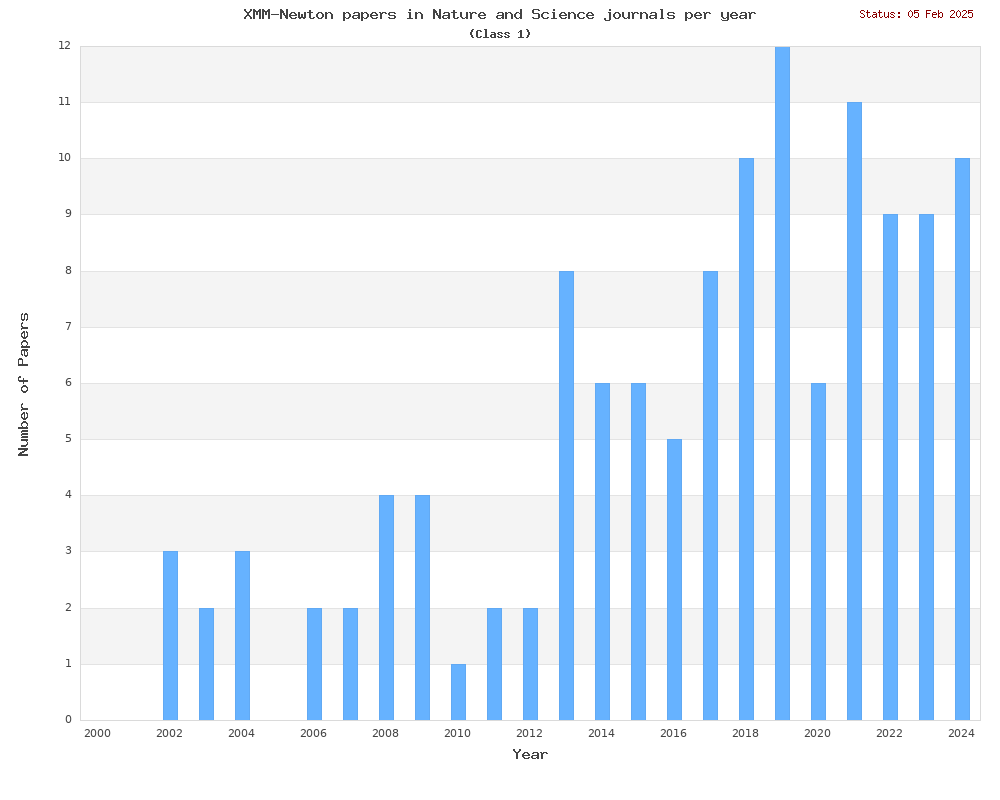}}
\caption{
XMM-Newton articles  published in the Nature and Science family of journals and  in  Class 1.
\label{fig6}}
\end{figure}

For the purposes of counting and citation, only classes 1 to 5 are considered, as they reflect the direct use of  XMM-Newton data or products, see Fig.~\ref{fig2}.
A bibliometric analysis of XMM-Newton articles has revealed that they are subject to elevated levels of citation. During the five years preceding this article, the mean percentage of articles that were in the top 10\%, 1\% or 0.1\% cited astrophysical articles was 31.5\%, 5.7\% and 1.2\%, respectively. 
The high-impact publications are illustrated by means of a visual representation of the class 1 papers published in the Nature and Science family of journals, see Fig.~\ref{fig6}.

\begin{figure}[t]
\centerline{\includegraphics[angle=0,scale=0.34]{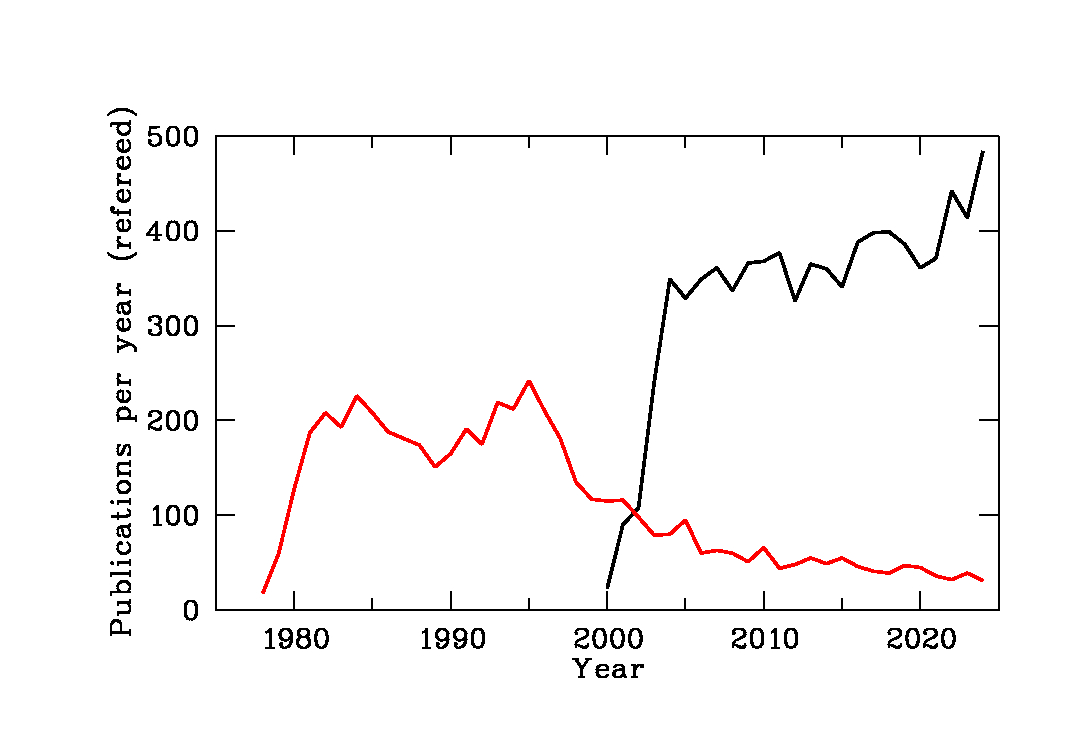}}
\caption{
Comparison of XMM-Newton publications  (Class 1 to 5, in black) alongside a similar analysis  for   IUE, in red. 
\label{fig3}}
\end{figure}

Figure~\ref{fig3}  illustrates the annual number of articles (class 1 to 5) observed for XMM-Newton compared to the International Ultraviolet Explorer (IUE). 
As an ultraviolet observatory, IUE
 provided the first astrophysical spectra for  this waveband, exhibiting observational characteristics comparable to those of XMM-Newton, albeit for a distinct wavelength range. 
The same classification schemes have been applied to the IUE articles as were used for XMM-Newton. 
Both curves demonstrate a pronounced increase during the initial five-year period. Subsequently, the IUE curve  declines from 1984 to 1990 and again from 1999 onward. Between 1990 and 1999, a peak is observed in the  IUE publication rate, probably  a result of joint programs with new facilities. The XMM-Newton  publication rate demonstrates a sustained growth trend from 2004 to 2024.

It is not feasible, due to limited resources, to quantify the contribution of XMM-Newton data to the conclusions of each article, as was done by \cite{2007AN....328..983T, 2008AN....329..632T,2010AN....331..338T}.
However, it is clear that, at the outset of the mission, the overwhelming majority of papers relied exclusively on XMM-Newton data. In contrast, approximately 30\% of articles currently employ XMM-Newton data to substantiate their conclusions.

A study of XMM-Newton articles was conducted by \cite{2014AN....335..210N,Ness2}, and the ensuing conclusions are outlined below. The number of articles published on an annual basis has now reached a total of more than 400 in 2022. These articles  allow us to characterise the community engaged in the research and publication of data obtained from XMM-Newton. 
On average, per year, approximately 130 researchers publish their first paper using XMM-Newton
data, and since 2010, a similar number of researchers has published
their final paper based on XMM-Newton data each year. Consequently, the total number of active researchers in this field has remained stable over time. 
The total size of the community is approximately 4,300 publishing scientists, of whom approximately 570 are first authors \citep{2014AN....335..210N, Ness2}. 
It should be noted that 75\% of the authors demonstrate sustained activity for a period of up to six years. Furthermore, 50\% of the authors publish only during one year, many publishing only a single paper \citep{2014AN....335..210N, Ness2}. This duration is typically observed for a PhD and postdoctoral fellowship. 
The remaining quarter of researchers engage with XMM-Newton data for a period exceeding six years, frequently publishing a paper on an annual basis.
The time elapsed between observation and the publication of the first article utilising the data exhibits a maximum at two years, with a potential secondary maximum at 3.25 years.
The scientific utilisation of XMM-Newton observations is exceedingly high, with more than 90\%  of the scientific time used in  refereed articles within a decade of the initial observation \citep{Ness2}.
It is also worthy of note that larger European optical facilities typically exhibit non-publication fractions of between 40 and 60\% of the observed programmes (see, for example, \citep{2017Msngr.170...51P}).

\begin{figure}[t]
\centerline{\includegraphics[angle=0,scale=0.33]{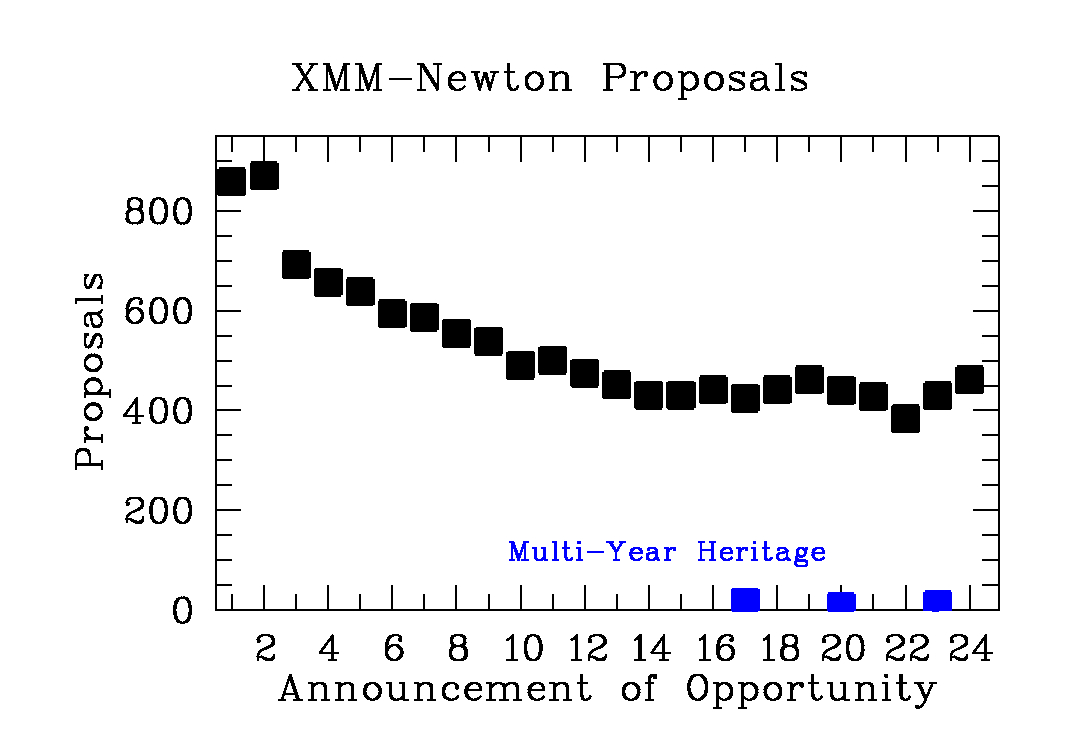}}
\caption{
 Number of XMM-Newton proposals received from AO-1 to AO-24. 
\label{number}}
\end{figure}

\begin{figure}[t]
\centerline{\includegraphics[angle=0,scale=0.33]{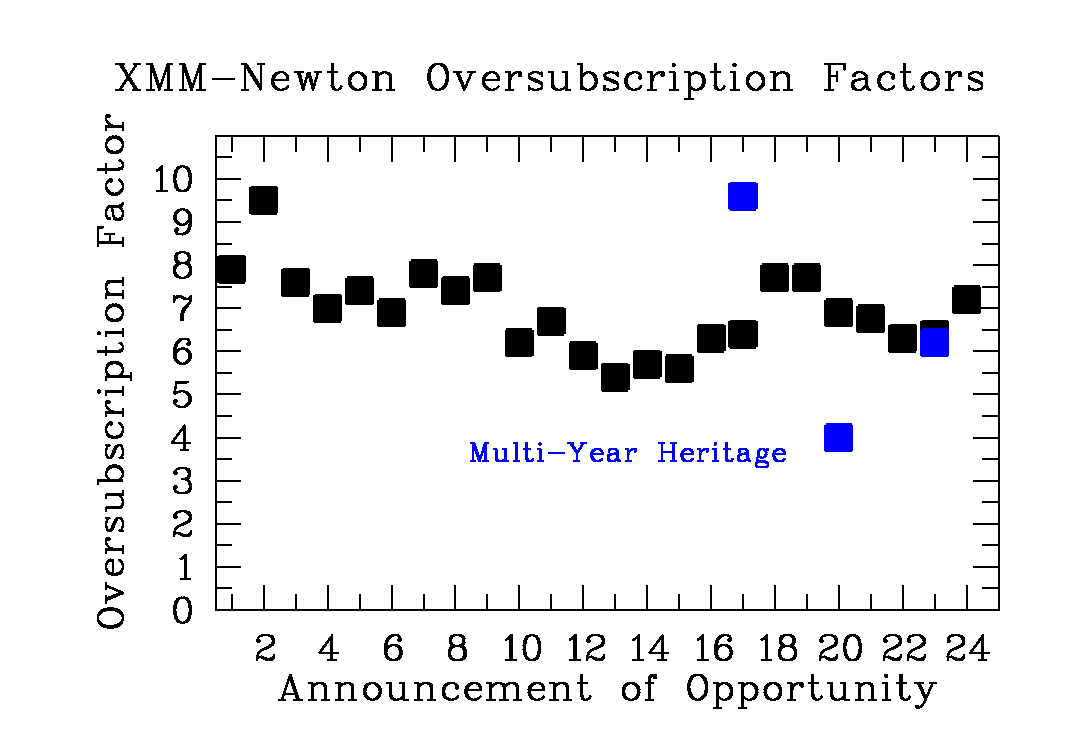}}
\caption{
Oversubscription factors for XMM-Newton AOs, with those for Multi-Year Heritage programmes indicated in blue. The lower oversubscription of these programmes in AO-20 is likely due to the COVID-19 pandemic. 
\label{over23}}
\end{figure}

The oversubscription of Announcements of Opportunity (AO) serves as an indicator of the scientific community's interest in pursuing further XMM-Newton observations, which in turn reflects the potential for scientific advancement.
Fig.~\ref{number}  illustrates the number of proposals received  from AO-1 to AO-24. Following the elevated figures observed in the initial two AOs (858 and 870 proposals, respectively), the number of received proposals declined until AO-12, at which point it stabilised at approximately 435. The blue symbols in the graph represent the number of Multi-Year-Heritage programmes received, with 16, 7 and 13 programmes, respectively.
Fig.~\ref{over23} provides a visual representation of the extent of oversubscription for XMM-Newton  AOs. 
Over the course of two decades, the oversubscription factors have remained consistently high, with a range of between 6 and 7 and consistently exceeding 5.5, 
with the exception of AO-2, which exhibited a factor of 9.6.

\section{XMM-Newton user model}\label{um}

The XMM-Newton user model is based on the premise of a modest research group, analogous to those typically found in European universities. 
All permanent members of staff are required to undertake significant teaching responsibilities, which leaves very little time for research at certain times of the academic year.  
Moreover, the research group may consist of non-permanent full-time researchers (e.g., postdoctoral fellows), postgraduate students, and other students, whose numbers may fluctuate and who may also have other responsibilities.  
It is recognised that the combined teaching and research environment, even in small research groups, can frequently give rise to new ideas and hypotheses, which may occasionally be of a transformative nature \citep{2025arXiv250910389M}, thereby leading to subsequent XMM-Newton programs. 
The distribution of years of activity as presented by \cite{Ness2}, with three quarters of the researchers being active for less than seven years, implies typical timescales for the master's, PhD and postdoctoral pathways, which supports  our assumed user model. 

In a considerable number of ESA Member States, the approval of an observation proposal for an ESA
mission serves as a guarantee of funding for postdoctoral or PhD positions. 
It is therefore evident that  successful proposals represent a crucial element in the attainment of funding. 
The relatively modest group size, combined with the presence of numerous non-permanent researchers whose tenure is limited in duration, results in a restricted pool of technical expertise available to the group.
It is therefore essential to focus on specific observing facilities and astrophysical object classes in order to ensure efficient research.
As the future of the research group is contingent upon the acquisition of new observing time, the extension of mission operations at every cycle represents a significant consideration that  requires strategic planning over  even longer timescales.

\section{Engagement of the Scientific Community}\label{com1}

The Science Strategy for the XMM-Newton Observatory represents the culmination of extensive consultations with the scientific community, particularly with those scientists who employ XMM-Newton data. 
Four principal channels of interaction between the scientific community and the Observatory facilitate the exchange of ideas and the formulation of new requirements:
(1) the XMM-Newton Users' Group (UG),
(2) the Observing Time Allocation Committee (OTAC), 
(3) conferences and workshops organised by the XMM-Newton Science Operations Centre, and
(4) the Helpdesk.

In 2001, the XMM-Newton UG assumed the role previously held by the Science Working Group. 
The change was prompted by a shift in focus from the construction of the spacecraft and its instruments towards the scientific operation of the mission.
Consequently, the primary audience was shifted from the instrument teams to the broader scientific community.
The XMM-Newton UG was chaired for a period of four years by Prof. Juergen Schmitt (Germany), Dr. Monique Arnaud (France), Prof. Xavier Barcons (Spain), Prof. Martin Ward (United Kingdom), Prof. Rudy Wijnands (Netherlands) and Dr. Anne Decourchelle (France).
The role of the chairs was of considerable consequence to the process. 
The XMM-Newton UG convened on an annual basis from 2004 onward. 
At the outset of the mission, in 2002 and 2003, the group convened on a biannual basis. The meetings serve as a platform for a comprehensive examination of the mission, encompassing its multifaceted aspects and operational elements. In particular, during the initial years of the mission, when XMM-Newton operations were adversely affected by understaffing and incomplete operational software, the UG played a pivotal role in facilitating communication with the scientific community. 

An annual  AO is issued by XMM-Newton, culminating in the final in-person meeting of the chairpersons of the different OTAC panels at the European Space Astronomy Centre (ESAC), where the SOC is located.  
The discussions held during this meeting, along with numerous contributions from panel members, have facilitated the optimisation and adaptation of the OTAC process to align with the evolving scientific and societal landscape.
The chairs of the OTAC have played a pivotal role in this process.

Since 2005, the SOC has organised a significant international conference on X-ray astrophysics, entitled "The X-ray Universe", on a triennial basis\footnote{see at https://www.cosmos.esa.int/web/xmm-newton/conferences}. 
Since 2011, the conferences have been held in European capitals with the objective of fostering links between the various national communities of ESA member states.
In the periods between conferences, the SOC has organised workshops$^1$ at ESAC, with a particular focus on  narrower themes within the field of high-energy astrophysics.
Conferences and workshops have consistently provided an effective forum for the exchange of ideas and the dissemination of knowledge regarding prospective avenues for advancement in the field.
In the context of the Science Strategy, the workshops held in 2007 \citep{2008AN....329..111S} and 2016 \citep{2017AN....338..139S, 2017AN....338..354S}, both entitled "XMM-Newton the next decade", proved instrumental in fostering dialogue and collaboration  within the scientific community, with a particular focus on exploring the scientific opportunities for the mission in the coming decade.

The Helpdesk offers comprehensive assistance to all scientists on matters pertaining to XMM-Newton, including operational aspects and scientific data analysis. Although the primary objective of the Helpdesk is not to engage with the scientific community, the queries it addresses have resulted in significant improvements in numerous areas of scientific assistance, particularly in the domains of calibration and documentation. 
The Helpdesk is run by the SOC at ESAC. 
There is another helpdesk facility at the GOF.

\section{The advancement of the scientific programme }\label{sec1}

\subsection{Discovery Space}\label{sec1}

On launch, XMM-Newton's substantial effective area, combined with its spatial resolution and spectral sensitivity, represented a very significant advance over what had earlier been possible.  While not an all-sky survey telescope (it only has a 0.2 square degree field of view), it can nevertheless study objects over an enormous range of distances, from the very nearest (within the Solar System) to the most distant galaxies and quasars.  It is important to recognise that the scientific ``discovery space'' of XMM-Newton is naturally limited by two factors: observational statistics and the distribution of potential targets in space.

All XMM-Newton's X-ray detectors are photon-counting, and hence subject to Poisson statistics.  This means that, when following up discoveries of new sources or new X-ray spectral features, the first step would be confirmation of detection or measuring line profiles, which would require higher signal-to-noise (S/N) data.  It would be normal to aim to double the S/N, but Poisson statistics tells us that would require quadrupling the number of counts i.e. quadrupling the observing time.  For efficient use of XMM-Newton, and for operational reasons, the typical observation time is 30 ks, with a minimum of 5 ks.  So, if these were the initial observations, then a follow-up with doubled S/N would require 120 ks.  And if further improvement was necessary then the next step would need 480 ks.  These requirements must then be considered relative to the typical time budget of an OTAC panel, which ranges from 350 to 850 ks for $\sim$30-40 competing proposals.  Consequently, the overall capacity to enhance S/N is constrained, as the ceiling observation times are reached very quickly.

\begin{figure}[t]
\centerline{\includegraphics[angle=0,scale=0.4]{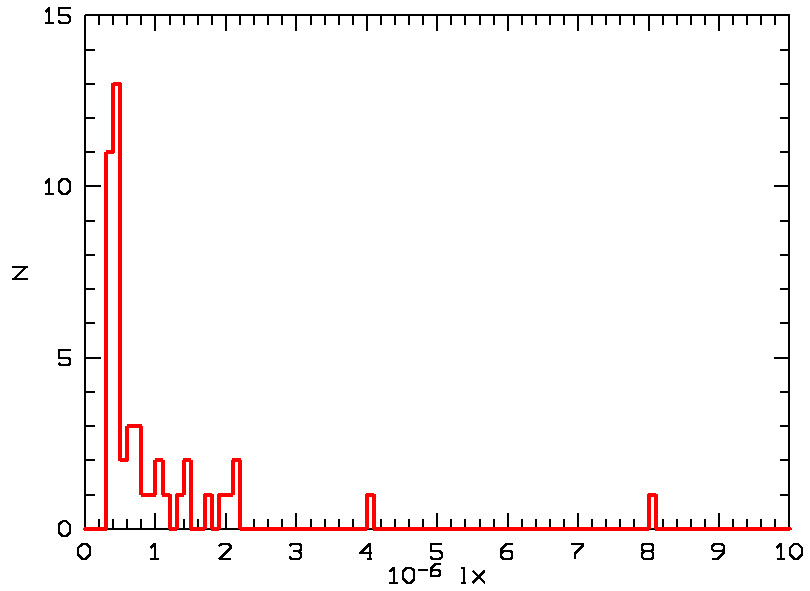}}
\caption{
Distribution of optical flux for the 30 brightest stars 
\label{star}}
\end{figure}

\begin{figure}[t]
\centerline{\includegraphics[angle=0,scale=0.4]{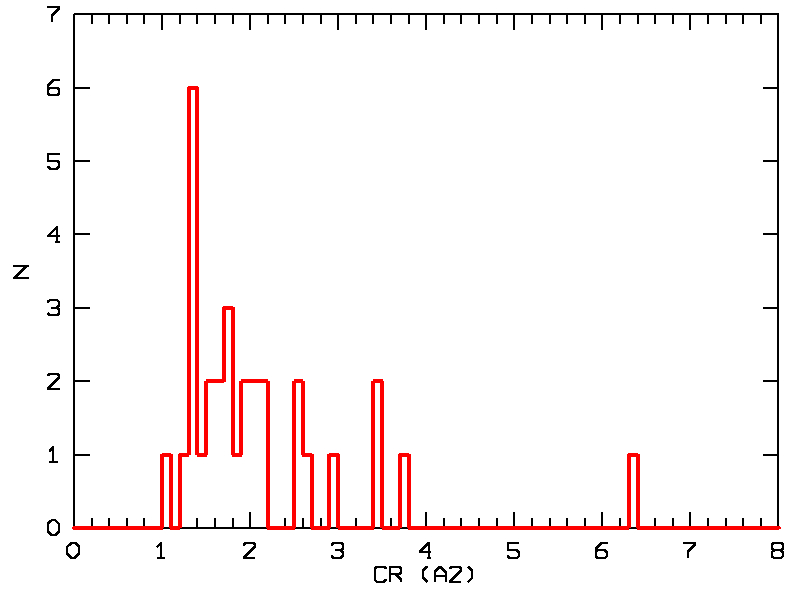}}
\caption{
Distribution of HEAO-1 A2 count rates for the AGN sample of \cite{1982ApJ...253..485P}   
\label{picc}
}
\end{figure}

An alternative  approach is to examine  X-ray properties across a range of sources, on the assumption that comparable celestial objects will exhibit analogous characteristics. It is illustrative to examine the optical flux of the 30 brightest stars, as illustrated in 
Fig.\ref{star}\footnote{\url{https://de.wikipedia.org/wiki/Liste_der_hellsten_Sterne}}.
 The brightest star is Sirius. The second brightest, Canopus, has a flux  half that of Sirius, while the third brightest star  is a factor of four less.  In fact,
that all stars with a flux below that of the 15 brightest stars are at least eight times lower.
Similarly, Fig.\ref{picc} depicts the X-ray count rate of the HEAO-1 A2 instrument  for the  \cite{1982ApJ...253..485P} Active Galactic Nuclei (AGN) sample. 
The Piccinotti sample is a complete, flux-limited sample of the 30 brightest AGN selected in the 2-10 keV band from the HEAO-1 A2 all-sky survey. 
One of the sources, NGC 4151, has a count rate of 6.34~{$s^{-1}$}. The second-highest count rate source, 3C 273,  is a factor of two lower.
AGN fainter than the Piccinotti sample of 30 AGN would exhibit a count rate lower than that of the brightest source by a factor of five or more,  as shown in Figure~\ref{picc}.
In general, the majority of object classes exhibit a paucity of bright sources, with only one or two exceptions. The remainder of the sources are below this level of brightness, with a few lying within a factor of five.

An exception to this behaviour is constituted by specific luminous, albeit very rare, galactic source types, such as outbursting black holes and neutron star binaries, which exhibit a distinctive spatial distribution.
However, no such sources are found within our immediate vicinity.
These sources are typically located in the central region of the Galaxy, at comparable distances and under comparable observational conditions.
An additional exception is  the Magellanic Clouds, where samples of rare but luminous galactic source classes, such as supernova remnants, can be studied, all essentially at the same distance.

A similar line of reasoning can be applied to population studies based on the observation of specific  sky regions. 
The number of found objects is proportional to the area surveyed, and follows a Bernoulli distribution, where  the error is proportional to the square root of the number.
In order to double the S/N for a given number of sources, it is necessary to quadruple the observed area, which in turn implies a factor of four increase in the observation time. 

With the exception of the flow of new discoveries, these constraints restrict the discovery space of a mission such as XMM-Newton. 
Typically, the known source classes are explored after a period of seven years, which 
 is approximately twice the duration  of a typical PhD thesis.
In a sense, this is a self-regulating time span, given that research in human-driven projects must be scaled to human time scales. 
To illustrate, it would take 21 years to double the S/N  of the data collected in seven years, which is incompatible with the timeframes typically allocated to research projects in  astrophysics. 

The construction of large observing facilities represents a significant financial commitment, and there is a clear incentive to ensure that they are utilised to their fullest potential.
Various strategies have been employed with the objective of extending the discovery space.
In the case of ground-based facilities, the installation of new instruments and upgrades frequently permits the investigation of new scientific topics. 
Given that the primary  mirror represents the most costly and defining component of the design, replacing and upgrading instruments frequently represents an economically viable strategy for opening up new research areas. 
Moreover, the rapid advancement of instrument technology in comparison to that of the primary mirror allows for regular replacement. 
Of all the astronomical satellite missions, Hubble is the only one designed to enable in-space servicing and instrument replacement.
NASA has demonstrated its capability to replace major components of  Hubble and its scientific instruments on five servicing missions, and the advantages of this have been emphasised by the scientific community \citep{2012AN....333..209S}. 
However, the design of XMM-Newton does not allow for this. 

In view of the aim of increasing utility, it is worth noting that a considerable number of
facilities owned by universities are repurposed for use as educational resources and development platforms for new instruments. 

An alternative strategy is to place the facilities in a survey mode, whereby extensive fields are observed in search of rare objects, such as quasars at the highest redshifts. 
The XMM-Newton orbit permits prolonged, uninterrupted observations; however, it also results in several hours during which the instruments are inoperable. Furthermore, XMM-Newton observations are frequently affected by elevated background radiation levels.
The grasp of view of XMM-Newton is of a very limited size, comparable to the 
 grasp of other space missions designed for doing large soft X-ray surveys, like ROSAT.
In light of the aforementioned constraints, it is evident that XMM-Newton is not an appropriate instrument for an all-sky survey. 

Another potential strategy would be to exploit the time domain through long-term monitoring. However, Sun, Earth, and Moon avoidance angles, together with the orbital and thermal constraints of XMM-Newton, limit the annual visibility of most sky regions to two periods of approximately two months, each followed by an interval of roughly four months during which the region is  inaccessible. Only a small fraction of the sky enjoys nearly continuous visibility throughout the year, while another small fraction is never visible at all. In addition, as described in sect. \ref{PeerReview}, the proposed observing programmes are evaluated and approved on an annual basis, which prevents the observatory from committing in advance to multi-year monitoring campaigns. As a result, long-term projects must be re-proposed each year, making them difficult to carry out and less likely to become sources of discovery. 
Nevertheless, some opportunities have arisen  through approved monitoring programmes or repeated observations of calibration fields. For 
example, the activity cycles of three Sun-like stars – HD 81809 \cite{2017A&A...605A..19O}, 61 Cyg,  and Alpha Centauri \cite{2012A&A...543A..84R} – were  successfully monitored by XMM-Newton over a 20-year period.

In order to maintain XMM-Newton's position at the vanguard of scientific research, the project has broadened the scope for discovery in three key areas: (1) Large, Very Large and Multi-Year Heritage Programmes, (2) co-ordinated observations with other observational facilities, and (3) targets of opportunity.

\subsection{Large, Very Large and Multi-Year  Heritage Programmes}
\label{Large}

In XMM-Newton's very first  AO, a request was made for 350 ks of observing time (A. Fabian: "A detailed study of the variable iron line in MCG–6-30-15", \cite{2002MNRAS.335L...1F}). However, this was considered to exceed the time budget allocated to the corresponding panel. 
Notwithstanding the absence of a mechanism for allocating time for such requests, the programme was approved at the final chairpersons' meeting.
The then Chairperson of the Time Allocation Committee, Prof. Malcolm Longair, underscored the scientific significance of approving such extensive time requests, as they allow the investigation of scientific questions that would otherwise remain unresolved.
Two long observing programmes, each comprising 650 ks and 500 ks of observation time, were approved in AO-2. 
The first of these was led by G. Hasinger and was entitled "Deep XMM spectroscopy in the Lockman Hole", \cite{2005A&A...441..417H}, while the second was led by J. Cottam,  "EXO 0748-6760", \cite{2008ApJ...672..504C}.
Both were subsequently accepted following deliberations at the chairpersons' meeting, with the latter made possible by the utilisation of Director's Discretionary Time (DDT)\footnote{The XMM-Newton Project Scientist (PS) is authorised to grant up to ~5\% of the total observing time outside the normal review process, which is referred to as DDT. }, thus augmenting the available time budget.
In AO-3, the formal introduction of large programmes was accompanied by the allocation of a dedicated budget for their implementation. The term "large programme" is employed to describe a project that necessitates a minimum of 300 ks of observation time. 
The evaluation of these programs follows a more rigorous approach, with a procedure specifically tailored to them as described in sect.~\ref{PeerReview} with the final decision being taken in the chairpersons' meeting.

The time allotted for large programmes is distributed in a manner that ensures they receive the same level of oversubscription as normal programmes for high-priority time, which is defined as time for which execution is guaranteed. 
For the majority of large programmes, the allocation of low-priority time, which is used as a filler, is not a viable option, given that the programme requires the full allocated time to achieve its scientific objective.
Two large programmes have been accepted in the COSMOS field (G. Hasinger, "Evolution of AGN in the cosmic web: the HST $+$VIMOS$+$SWIRE COSMOS field", \cite{2007ApJS..172...29H}) in AO-3 and AO-4, with a total of 1.27 Ms, representing an exceptionally long exposure time for an early AO.
In AO-4, a further large programme was conducted utilising DDT (R. Mushotzky: "SWIFT AGNs", \cite{2008ApJ...674..686W}, 300 ks).
From this point onwards, the total time allocation was sufficiently flexible to obviate the necessity for DDT.

In AO-7, a programme of 3 Ms of observing time was proposed (A. Comastri: "The ultradeep survey in the CDFS: an XMM-Newton legacy", \cite{2011A&A...526L...9C}). Of this, 1.5 Ms were accepted, with the recommendation that the remaining time should be proposed in the next AO cycle.
The remaining time was approved in AO-8 following contentious discussions at the chairpersons meeting.
Approval on a partial or two-stage basis presents a number of challenges, including the following: Approximately half of the chairpersons are new and therefore did not participate in the previous discussion and decision-making process. 
The number of observations conducted so far may be insufficient to substantiate the findings.
 Furthermore, the page limit for the scientific rationale would have to pertain to the observation, which would detract from the original scientific justification.

\begin{table*}[]
    \centering
    \title{Very Large and Multi-Year-Heritage Programms}
    \begin{tabular}{rllll}  \hline \hline
          AO & Type & Time & Principal Investigator  & Title \\  
             &  & [ks] &   & \\  \hline 
          7  & VLP & 3000 & A. Comastri & The ultradeep survey in the CDFS: an XMM-Newton legacy \\
             &  &  &  &  {\citep{2011A&A...526L...9C}} \\
          8 & VLP & 2900 & M. Pierre & The ultimate XMM  extragalactic survey {\citep{2016A&A...592A...1P}} \\
         11 & VLP & 1900 & F. Haberl & A survey of the Large Magellanic Cloud {\citep{2017A&A...598A..69H}}  \\
         12 & VLP &1628 & J. de Plaa & The XMM-Newton view of chemical enrichment in bright galaxy  \\
             &  &  &  &  clusters and groups {\citep{2017A&A...607A..98D}}\\
         13 & VLP & 1209 & D. Eckert & The XMM-Newton Cluster Outskirts Project {\citep{2017AN....338..293E}}  \\
         14 & VLP & 1626\footnote{The pregame was redefined in AO15 due to a significant change in the source flux} & F. Nicastro & Securing the Detection of the Hot Roaming Baryons with the \\
         &  &  &  & XMM-Newton {\citep{2018Natur.558..406N}} \\
         14 & VLP & 1379 & A. Boyarsky & Probing the Dark Matter Nature of the 3.5 keV Line with Draco \\
             &  &  &  &  {\citep{2016MNRAS.460.1390R}} \\
         15 & VLP & 1224 & W.N. Brandt &  Going Beyond COSMOS with the XMM-SERVS Survey of   \\
         &  &  &  &  W-CDF-S, XMM-LSS, and ELAIS-S1 {\citep{2018MNRAS.478.2132C}} \\
         15 & VLP & 1548 & A.C. Fabian & Mapping the inner accretion flow: dynamic
          reverberation in   \\
         &  &  &  &  IRAS 13224-3809 {\citep{2017Natur.543...83P, 2020NatAs...4..597A}} \\
        16 & VLP & 1045  & G. Risaliti  &   Cosmology with z$>$3 quasars 30 {\citep{2019A&A...632A.109N}}  \\  
         17 &  MYH & 3400 & W.N. Brandt  &   Completing and Ensuring Major Impact from the XMM-SERVS \\
         &  &  &  &          Survey {\citep{2021ApJS..256...21N}}    \\ 
         17 & MYH & 3000  & M. Arnaud and S. Ettori  &   Witnessing the culmination of structure formation in the Universe \\
             &  &  &  & {\citep{2021A&A...650A.104C}}    \\
          20 & MYH & 3600  &  G. Ponti  &   Is the activity in the Milky Way disc sustaining the Galactic corona?\\
             &  &  &  &  {\citep{2024A&A...686A.125M}}   \\ 
          20 & MYH & 2400 & L.  Zappacosta  &   HYPerluminous quasars at the Epoch of ReionizatION \\
             &  &  &  & {\citep{2023A&A...678A.201Z}} \\
          23 & MYH & 3576 &  M. Pierre, S. , Maughan   & X-EDFF \cite{}   \\  
           &  &      & and M. Bolzonella  &    \\  
          23 & MYH & 2331 & G. Lanzuisi   &  The WISSHFUL Program: unveiling SMBH winds at cosmic noon
          %\cite{}
          \\     \hline  
    \end{tabular}
    \caption{ Very Large (VLP) and Multi-Year Heritage (MYH) Programmes
    \label{VLPMYHP} }
\end{table*}

In order to address this issue, the introduction of Very Large Programmes was deemed necessary, with an observation time of 1 Ms or more.
Such programmes were treated as large programmes, but were discussed separately at the chairpersons' meeting. 
In the case of programmes of considerable size, OTAC was able to allocate up to 1.5 Ms for observation in the subsequent AO. 
This approach enabled the accommodation of substantial requests while also addressing visibility constraints, given that a considerable portion of the sky is only visible for 12 weeks annually.
Tab.\ref{VLPMYHP}  presents a list of the accepted Very Large Programmes, indicating the approved observing time, the Principal Investigator, and the title.

In AO-15, N. Brandt proposed a programme \citep{2015xmm..prop...20B, 2018MNRAS.478.2132C} with an observing time of 4.6 Ms, with the objective of observing three extra-galactic fields in addition to the existing Cosmos field.
The objective was to mitigate the effects of cosmic variance, which was evident in the data obtained from the COSMOS field. 
OTAC accepted only one of the three fields, a decision that recalls the circumstances of AO-7, which resulted in the introduction of Very Large Programs. 
In AO-16, no additional fields were approved, reflecting the inherent dilemma that the full time required to achieve the scientific goal could not be approved in a single AO. This indicates that the realisation of the scientific objective is contingent upon future decisions by OTACs, the specifics of which are currently unknown.

Consequently, the Very Large Programmes have been superseded by Multi-Year Heritage Programmes. 
In order to be considered for inclusion in the Multi-Year Heritage Programmes, applicants are required to request a minimum of 2Ms of observing time. 
The programmes are convened triennially, with up to 6~Ms of observing time available, and to be observed over a three-year period.
The selection process is overseen by a special panel, the Senior Review Panel, see Section~\ref{PeerReview}

\begin{figure}[ht]
\centerline{\includegraphics[angle=0,scale=0.29]{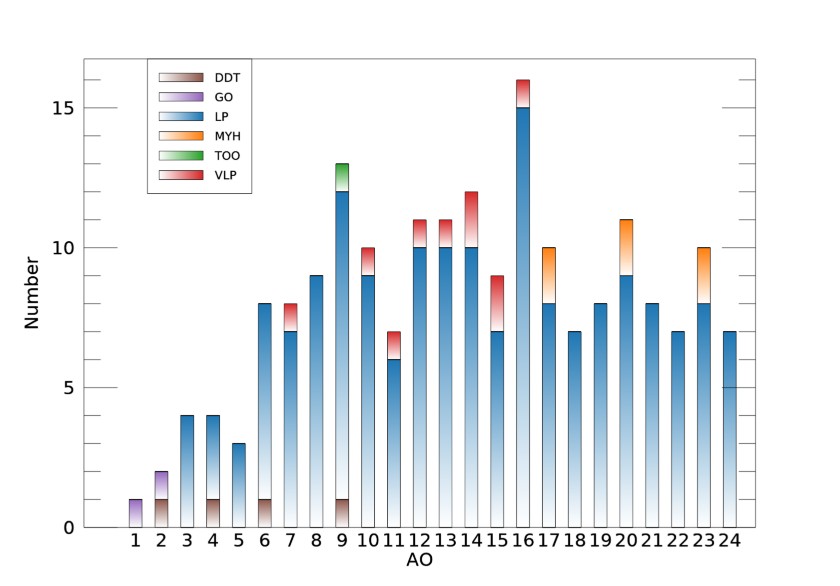}}
\centerline{\includegraphics[angle=0,scale=1.25]{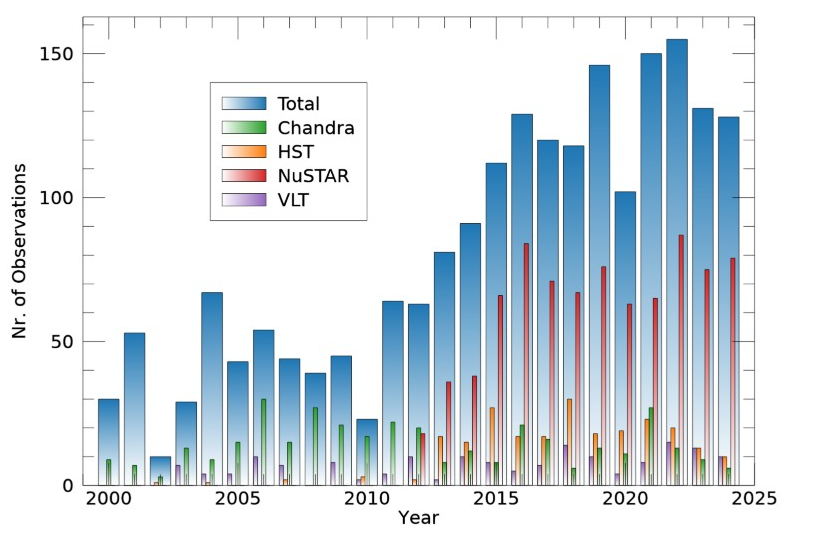}}
\centerline{\includegraphics[angle=0,scale=0.29]{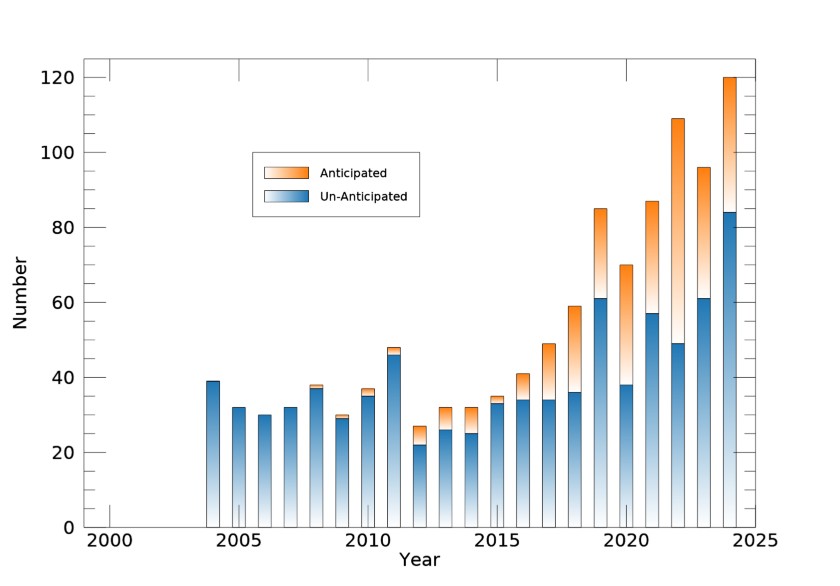}}
\caption{Evolution of the XMM-Newton science programme: (Top) programmes with duration $>$300 ks; (Middle) number of coordinated observations; (Bottom) number of Target-of-Opportunity observations.
 \label{EvoNum}}
\end{figure}
\begin{figure}[ht]
\centerline{\includegraphics[angle=0,scale=0.29]{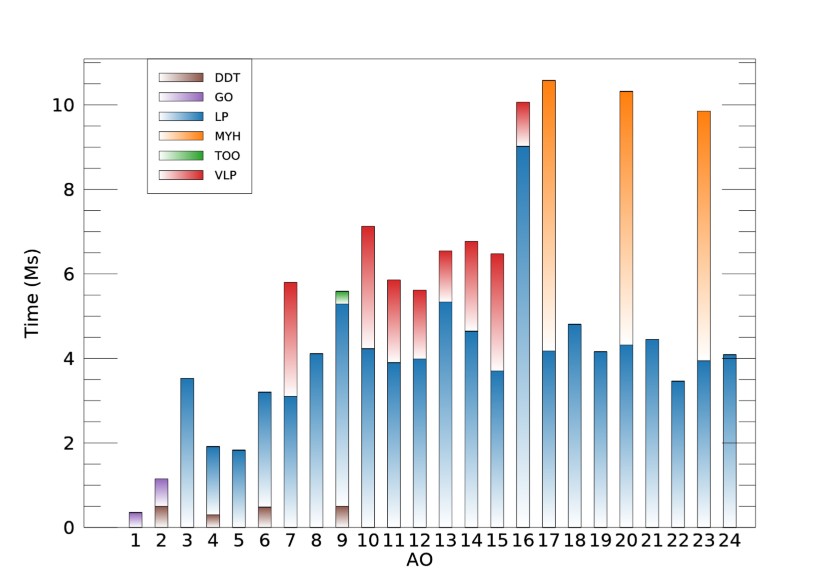}}
\centerline{\includegraphics[angle=0,scale=1.25]{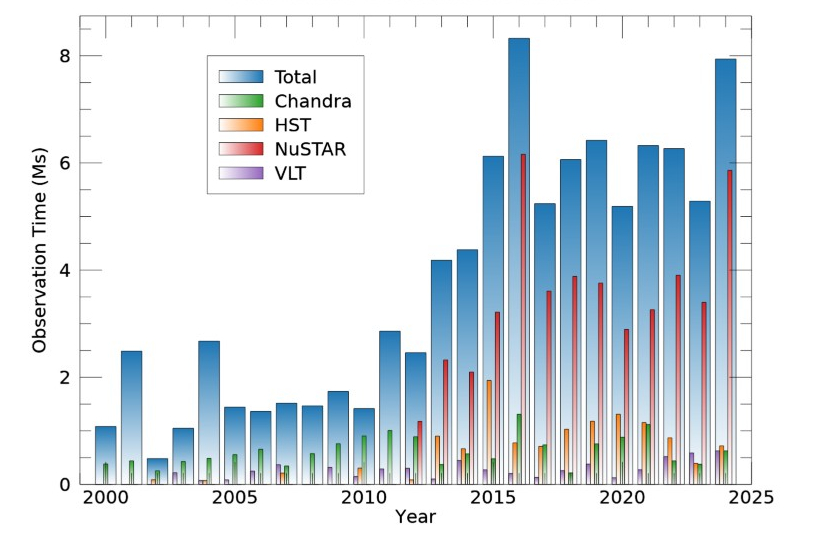}}
\centerline{\includegraphics[angle=0,scale=0.29]{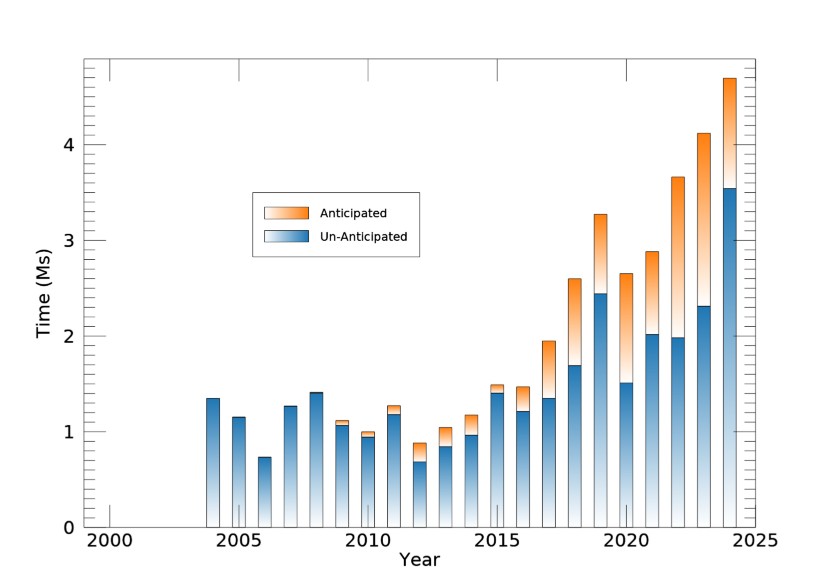}}
\caption{ Same as Figure~\ref{EvoNum}, but for observing time. \label{EvoTime} }
\end{figure}

Figure \ref{EvoNum} depicts the number of programmes accepted in each AO with a total exposure time exceeding 300 ks, while Figure \ref{EvoTime} (top panel) illustrates the total observing time of the programmes accepted in each AO with a total exposure time exceeding 300 ks.
In the initial stages of the observing cycle, up to and including AO-5, the number is less than five. Subsequently, from AO-6 to AO-9, the number of accepted programmes increased to approximately thirteen,
and has now stabilized at around nine, corresponding to around six million seconds of observing time per AO.

The advent of the Large Programmes, followed by the introduction of the Very Large Programmes and the Multi-Year Heritage Programmes, was precipitated by requests for comprehensive or extensive extra-galactic field observations.
In each instance, the requested timeframe exceeded the flexibility permitted by the established proposal and time allocation rules. 
Thus, in each case, the regulations were either introduced or redefined with the objective of facilitating the allocation of time in subsequent calls.
It is worthy of note that following the initiation of extra-galactic surveys, a galactic survey was subsequently approved following the implementation of the Very Large Programmes and the introduction of the Multi-Year Heritage Programmes.
(F. Haberl: "A survey of the Large Magellanic Cloud", \citep{2017A&A...598A..69H} (1.9 Ms) and G. Ponti: "Is the activity in the Milky Way disk sustaining the Galactic corona?" \citep{2024A&A...686A.125M} (3.6 Ms).

Since AO-7, approximately 55\% of the high-priority observing time (A and B priority\footnote{The execution of A and B priority observations is guaranteed whereas C priority observations are used as fillers and only a part of them is observed.}) has been allocated to Large and Very Large Programmes, as well as subsequent Multi-Year Heritage Programmes.
This equates to a mere 10 programmes per AO. 
Principal Investigators (PI)s, as well as members of the OTAC, are confronted with the considerable challenge of the high level of oversubscription for XMM-Newton observing time, which is typically by a factor of six. 
Given the restricted number of approved large programmes, the observational approach is exploratory in nature, whereby the experiment is conducted on a single occasion and solely for a single target. In the majority of cases, the time and resources available are insufficient to conduct a second observation with the same scientific objective for a second source. In order to conduct a replication of an experiment, it is necessary that the source in question possesses characteristics that allow for the drawing of new conclusions, despite the confirmation of the effect in question.

 A key feature of this process was to apportion the available time for large programmes in a manner that ensures the same level of oversubscription  applies to large programmes as to normal programmes.
This resulted in 55\% of the allocated observing time being dedicated to large programmes, thereby facilitating the exploration of new areas.
In comparison to the typical observing time of 30 ks, the newly allocated 3 Ms permits an increase in exposure time by a factor of 100, thereby enhancing the  S/N by a factor of 10. 

The exploratory nature of the large programmes is reflected in the recent distribution of Very Large and Multi-Year Heritage programmes from AO-16 onwards, as illustrated in Table~\ref{VLPMYHP}.
On the one hand, there are galactic and extra-galactic surveys that are much larger than ever before; on the other hand, there are three programmes targeting quasars at high and highest redshifts.
(G. Risaliti: "Cosmology with z$>$3 quasars" {\citep{2019A&A...632A.109N}} (1.045 Ms); L. Zappacosta: "Hyperluminous quasars in the epoch of reionisation" {\citep{2023A&A...678A.201Z}}  (2.4 Ms); G. Lanzuisi: "The WISSHFUL Program: unvealing the SMBH winds at cosmic noon" (2.3 Ms).

\subsection{Coordinated Observations and Joint Programmes}

The combination of observations from two or more facilities 
permits the exploitation of their complementary capabilities as well as the extension of the spectral  range covered, thereby increasing the available discovery space.
In many cases, the combined dataset enables the resolution of scientific questions that would not be possible  from a single facility.
From the first AO onwards, the concept of ``added value'', which generally refers to supplementary observations conducted with other facilities, has been viewed favourably in the peer review of proposals.
As it is typically challenging to request observing time in two distinct review processes, XMM-Newton permits highly rated observations to be postponed up to two times into the subsequent observing cycle, thereby facilitating joint observations.
At the end of 2024, 2252 XMM-Newton observations  have been coordinated with 76 different facilities, including virtually every major astrophysical observing facility.

Figure \ref{EvoNum} (middle panel) illustrates the total yearly number of XMM-Newton observations that have been coordinated with other facilities. Furthermore, the figure illustrates the subsamples of coordinated observations with Chandra, HST, NuSTAR, and VLT, which are represented by different colours. 
Figure \ref{EvoTime} (middle panel) illustrates the same data in terms of observation time. From 2000 to 2010, the number of coordinated observations was less than 50 per annum, with an equivalent observing time of approximately 1 Ms. From 2011 to 2016, there was a notable increase in the number of coordinated observations, reaching approximately 130 (6 Ms) per annum. Thereafter, the number remained relatively constant.

It should be noted that these figures represent lower bounds for  coordinated programmes, as there are numerous additional coordinated observations organised by the PI subsequent to the XMM-Newton schedule becoming available. These observations are not reflected in the SOC data set. 

In order to facilitate coordinated observations involving a considerable number of requests, joint programmes were established. 
The inaugural joint programme of XMM-Newton was established in collaboration with Chandra, following a suggestion put forth by the SPC.
The suggestion was made following a discussion about the relatively short time distance between the launch dates of the two missions, in comparison to the original planning, which had foreseen that the two missions would follow each other with an interval of years. 
As XMM-Newton is classed as a "large" astronomical facility in the United States, it is subject to the United States policy of connecting "large" astronomical facilities with joint programmes. Such entities include the National Radio Astronomy Observatory (NRAO), the James Webb Space Telescope (JWST), the Hubble Space Telescope (HST), and the Chandra X-ray Observatory in the case of XMM-Newton. 
A second joint programme was established with ESO VLT/VLTI, which serves as the principal European optical/NIR facility.
This program was offered continuously from AO-4 onward  after a test phase in AO-2. 
XMM-Newton also engaged in joint programmes with the ESA observatories INTEGRAL and Herschel.
The co-location of the science operations centres of XMM-Newton and INTEGRAL, especially for TOOs, permitted the scheduling of a considerable number of additional simultaneous observations, which held the potential for added value.
XMM-Newton  entered into an agreement with the Neil Gehrels Swift, whereby the XMM-Newton peer review process can allocate time on Swift for observations.
This allows for the undertaking of brief Swift observations with the objective of monitoring a source or searching for specific states.

At the time of writing, the most fruitful joint programme is with NuSTAR. 
The OTAC accepted a considerable number of requests for simultaneous XMM-Newton observations of NuSTAR performance verification observations prior to the launch of the mission.
This was in contradiction with the prevailing view that approval of targets for simultaneous observation should be withheld until the performance of the other, new, facility had been demonstrated.
In light of the favourable outcome of the proposals, the project scientists of both missions reached a consensus on the implementation of a joint programme with a budget of 1.5 Ms for each mission resulting in 3 Ms of joint observing time per year.
Ultimately, joint programmes with MAGIC and HESS were established, reflecting the high utilisation of XMM-Newton observations in conjunction with TeV detection.

Furthermore, the majority of coordinated programmes require simultaneous or quasi-simultaneous observations. The term "quasi" indicates that the interval between the two observations is relatively brief in comparison to the anticipated variability of the source. 
Joint programmes for Planck-detected clusters of galaxies provide an illustrative example of a programme where the timing of the observations was not a significant factor. In this programme, the higher spatial resolution of Chandra was employed to resolve the cluster centre, whereas the larger field of view of XMM-Newton permitted  observation of the entire cluster.

The joint programmes of XMM-Newton encompass a broad range of energies, extending from radio to TeV.
It is rare for a single physical emission mechanism to account for a broad energy range.  Nevertheless, there are notable exceptions to this, including gamma-ray bursts (GRB) and the synchrotron emission of blazars. 
In the majority of cases, different energy ranges are dominated by disparate emission mechanisms or even by different sources. 
In the case of clusters of galaxies, the optical light is dominated by the stars in the individual galaxies, whereas the X-rays are emitted by the hot intercluster gas.
In the case of galaxies and  AGN, the application of spectral energy distribution (SED) fitting based on a range of optical to infrared filters enables the  constituent components to be identified, thereby facilitating the  redshift to be estimated.

The success of a joint programme can be attributed to two key factors: (i)  the extent to which the observing parameters of the respective facilities are aligned; (ii)  the extent to which the two facilities are complementary for the study of different physical processes. 

Both INTEGRAL and NuSTAR permit the extension of  XMM-Newton's energy range towards higher energies, which is of significant importance for  many studies. 
INTEGRAL utilises a coded mask instrument, which constrains both its spatial resolution and sensitivity.
To illustrate, the majority of AGN observed by XMM-Newton are too faint to obtain a spectrum with INTEGRAL. Conversely, a considerable number of typical INTEGRAL  sources, such as bursting Galactic black hole binaries, have a brightness that exceeds the capabilities of the XMM-Newton detectors. 
The spatial resolution and sensitivity of NuSTAR are markedly superior to those of INTEGRAL, thereby facilitating the observation of a greater number of sources suitable for spectroscopic studies that are accessible to XMM-Newton.

The rationale behind joint observations is that the same physical mechanism can be constrained in different ways by both facilities.
The principal motivations behind numerous joint XMM-Newton  and  HST programmes are the impact of outflows, winds and absorption systems at varying ionisation levels of AGN and the interconnection between disparate systems. 
XMM-Newton/NuSTAR observations represent the most frequently requested and performed joint programmes.
The joint use of both instruments permits the characterisation of the 6.4 keV iron line complex, including the line and continuum emission, as well as the identification of absorbing components.
The reverberation mapping of compact objects, which enables the measurement of supermassive black hole mass, spin and coronal height, represents the most impressive example of this undertaking \citep{2020NatAs...4..597A}.
Furthermore, a formal joint programme did not exist for XMM-Newton and Planck, yet both were able to address the hot gas in clusters of galaxies. This resulted in a series of X-ray follow-up observations of clusters of galaxies detected by Planck {\citep{2021A&A...650A.104C}}.
XMM-Newton is able to directly measure the thermal emission from the hot intracluster gas, whereas Planck measures the imprint of this gas on the microwave background via the Sunyaev-Zel'dovich effect.

The identification of new HESS and MAGIC-detected TeV sources requires the presence of counterparts at other wavelengths. In the majority of cases, XMM-Newton is the optimal instrument for this purpose. To illustrate, supernova wind nebulae, which display bright X-ray sources that are barely discernible in the optical domain, are particularly well-suited for observation with XMM-Newton. 
In contrast, the joint Herschel/XMM-Newton programme yielded a single accepted proposal, namely an observational study of SgA*. It is well established that a significant number of SgA* flares are detected in the X-ray and infrared energy ranges \citep{2015MNRAS.454.1525P, 2017A&A...604A..85M}, which are likely caused by synchrotron emission \citep{2017MNRAS.468.2447P}. Nevertheless, the majority of infrared sources do not exhibit a correlation with X-ray counterparts.

\subsection{Targets of Opportunity}

Targets of Opportunity (TOOs) are defined as astronomical events observable by XMM-Newton which cannot be predicted or scheduled on a yearly timeframe. Nevertheless, they are scientifically significant enough to justify the interruption of the scheduled programme.
These TOOs are further categorised into two distinct classifications: those that are accepted within the AO and designated as Anticipated TOOs, and those that are wholly unanticipated and termed Unanticipated TOOs.

The observation of  TOOs was not a major factor in the design of the XMM-Newton spacecraft, its instruments, the onboard and ground computer, and the software.
In order to address this challenge, investigators involved in AO-1 were permitted to propose triggered observations, which enabled the observation of known sources in specific states, at a minimum.
The concept of  TOOs was only introduced in operations following the approval of the ESA Director of Science in response to a question posed by a member of the public during a press conference.
However, failing to strongly consider TOOs  during the design phase imposes constraints on the technical ability to perform observations with a short reaction time and necessitates a significant manual workload.  
For instance, observations conducted with the optical monitor are only permitted to be terminated once the current exposure period (which lasts between 2 and 25 ks) has concluded. This is to guarantee the safety of the instrument. 
The development of software for handling TOO alerts and the subsequent  observations facilitates the management of TOOs. Nevertheless, this necessitates manual intervention, which would be considerably reduced in an operational system where TOOs are integrated from the outset of the design phase onwards.

Figures ~\ref{EvoNum} and ~\ref{EvoTime} illustrate the number and time spent, respectively, of TOO observations conducted over time, with colours indicating both anticipated and unanticipated TOOs for each year. 
In the initial phase of operations, all TOOs were conducted in Director's Discretionary Time (DDT). 
Up to 2016, approximately 35 TOO observations were conducted annually, equating to approximately 1 Ms of observing time.
 Subsequently, the figure rose to 110 per year in 2023, implying an observing time of 4 Ms.

From AO-7 onwards, investigators were permitted to submit proposals for anticipated TOOs as part of the call for observing time proposals.
In A0-8, there was a possibility that the allocated time for DDT would be insufficient to meet the requests for unanticipated TOOs.  In order to ensure the feasibility of substantial unanticipated TOO observations, Prof. Brian McBreen, the chairperson of OTAC, called for a solution that would prevent the limited allowable allocation of discretionary time  from impeding unanticipated observations.
From AO-9 onwards, when possible, unanticipated TOOs have been sent to an OTAC panel chairperson for review, and those recommended are no longer included in the DDT account.
This established a  higher degree of flexibility in the allocation of time for unanticipated TOOs and has permitted an increase in the number of unanticipated TOOs from 2015 onwards.

The process of time allocation does not impose any restrictions on the number of accepted TOOs. 
The majority of panels accept only a single TOO proposal per object class.
 However, it should be noted that up to four panels may accept requests for the same object class, which may result in the acceptance of multiple TOOs per class. 
In addition, proposals for TOOs are valid for a period of three cycles\footnote{The acceptance of TOOs is contingent upon the allocation of a high observing priority, which consequently enables the observation to be shifted up to two times to the subsequent AO}. This facilitates coordination with other facilities, but has the potential to result in the establishment of further multiple TOOs. 
It is also worth noting that, in some cases, an event that triggers several anticipated TOOs may also lead to unanticipated TOOs.
 Moreover, the scientific classes are not entirely distinct, with some degree of overlap. To illustrate this point, it is expected that a neutron star–neutron star merger will occur as both a short gamma-ray burst (GRB) and a gravitational wave (GW) event, which implies the existence of different triggers for the same event.

A further challenge arises from the fact that different facilities may identify the same event as a trigger. To illustrate, a bright gamma-ray burst (GRB) discovered by INTEGRAL and a bright GRB identified by Swift may both be indicative of the same event.
In view of the considerable number of anticipated triggers that are concurrent, XMM-Newton TOOs are typically not exclusive, thereby placing the responsibility on PIs to trigger promptly. From the standpoint of the project as a whole, there are two advantages. Firstly, the system has redundancy in the event that a scientist is unable to trigger the corresponding proposal, as there are generally other proposals triggered. Secondly, no scientist is able to exercise a {\it de facto} monopoly over a specific area of scientific inquiry, which is particularly advantageous in comparison with missions that have limited TOO budgets. 
A potential disadvantage is that, in the case of prominent events and equivalent observing strategies, the PS must decide which proposal to observe. In doing so, the PS considers the AO of proposal acceptance, where the  oldest has priority, and the time the actual trigger was received.

The ability of XMM-Newton to identify novel transient phenomena that are sufficiently luminous to warrant subsequent observation and investigation is limited.
In order to identify new and rare transient sources, it is necessary to conduct a comprehensive survey of a significant portion of the sky.
The utilisation of long XMM-Newton slews has facilitated the scanning of regions within the celestial sphere with brief exposure times of a few seconds. This approach has led to the discovery and publication of one Quasi-Periodic Eruption (QPE) \citep{2013MNRAS.433.1764M, 2019Natur.573..381M} and one QPE candidate \citep{2021ApJ...921L..40C}, seven Tidal Disruption Events (TDE) \citep{2007A&A...462L..49E, 2008A&A...489..543E, 2012A&A...541A.106S, 2017A&A...598A..29S, 2019A&A...630A..98S, 2020ApJ...891..121L}, one flaring AGN \citep{2016MNRAS.461.1927P}, one supernova \citep{2017ApJ...850..111N}, two novae \citep{2008A&A...482L...1R, 2009A&A...506.1309R} and one RS CVn variable \citep{2012PASP..124..682L}. 

The XMM-Newton pointings have facilitated the detection of hitherto undetected transient sources, exhibiting significant scientific merit.
This can be illustrated by findings of the SSC: five total or partial TDE \citep{2011ApJ...738...52L, 2015ApJ...811...43L, 2017NatAs...1E..33L, 2018NatAs...2..656L, 2009Natur.460...73F}, one QPO \citep{2023A&A...675A.152Q}, one TDE showing QPO \citep{2013ApJ...776L..10L, 2017MNRAS.468..783L}, one new and one   pulsating ULX \citep{2013ApJ...779..149L, 2021MNRAS.503.5485Q}, one unusual pulsar \citep{2017ApJ...839..125Z},  two cataclysmic variables \citep{2018A&A...615A.133W, 2020A&A...637A..35S} and one extragalactic Type 1 burst \citep{2020A&A...640A.124P}.  Nevertheless, the constrained field of view imposes limitations on the available discovery space. Given its substantial effective area, XMM-Newton allows the identification of the faintest transients. However their identification in other wavelength bands is challenging, and the X-ray spectral analysis of these transients is constrained by the limited number of counts collected.

The majority of transient objects are identified by facilities with a larger field of view, sky monitors and, recently, projects that systematically observe repeatedly a significant portion of the sky.
The initial unanticipated transients were found to be dominated by gamma-ray bursts (GRBs), which BeppoSAX was the first to detect with a positional accuracy sufficient to permit subsequent observations \citep{1997IAUC.6787....1H}.
Following the launch of Swift, XMM-Newton observations were increasingly restricted to the brightest GRB events and to special GRB cases that were likely caused by new source classes.
A pertinent example is the detection of an outburst of a magnetar by gamma-ray instruments and XMM-Newton TOO observations during the decline phase.
At the present time, sky monitoring surveys, such as Zwicky transient facility (ZTF) and ATLAS Transient Science Server, are responsible for the discovery of the majority of transient sources associated with XMM-Newton TOOs.

Over the past 25 years, the source types observed as TOO have demonstrated a significant degree of evolution, reflecting  advances in our comprehension of the various source classes.
 TDEs were exceedingly uncommon, with a considerable proportion of them identified during XMM-Newton slews.
From approximately 2020 onwards, ZTF began to regularly detect new TDE in the optical, which were subsequently followed by X-ray emission over a months-long baseline.
One of the key advantages of ZTF is the relatively short time interval between observations, which allows for a clear determination of the occurrence date.
Presently, XMM-Newton TOOs of TDE are investigating a range of disparate behaviours (rebrightening, radio emission, QPE), whereas observations of "canonical" TDE do not justify further investment of observing time.

The Einstein Probe \citep{2022hxga.book...86Y} 
mission is currently at the forefront of a novel  approach, conducting a comprehensive survey of a vast area in the X-ray domain with the objective of identifying transient X-ray sources, including TDE.
It is anticipated that this will be succeeded by the Ultraviolet Transient Astronomy Satellite \citep{2024ApJ...964...74S},
UltraSAt, which is scheduled to take place in the fourth quarter of 2027. 
In addition, the Space-based multi-band astronomical Variable Objects Monitor \citep{2023hxga.book..149W}, 
SVOM, has been searching for and characterizing new transients at gamma-ray energies since 2024.

The TOO requests are subject to a certain degree of self-regulation.
As a result of each additional TOO observation, the classification and understanding  of source classes are becoming increasingly precise. Consequently, the justification for further observations of a particular source type likely becomes increasingly tenuous, particularly in light of the availability of a representative sample. 

\subsection{Incorporation of new communities}

The advent of new scientific knowledge frequently gives rise to the formation of new research communities. It is therefore essential to guarantee the continued accessibility of XMM-Newton on the widest possible scale. It is of the utmost importance that comprehensive documentation, an extensive archive, sophisticated analysis software and readily available data products be provided, as these elements serve as fundamental pillars, facilitating straightforward access for scientists lacking expertise in X-ray astronomy and specifically XMM-Newton.

A fundamental aspect of the scientific process is the formulation of hypotheses, which are then subjected to peer review and subsequent observation, analysis and publication.
It is only through the submission of a proposal that one can present novel and original ideas for consideration. 
Although the scientific rationale permits a wide range of argument, the feasibility study represents a crucial juncture in the process. It is at this stage that the panel must be persuaded of the feasibility of the scientific study, based on the desired observations.

The process of assisting new communities in becoming integrated into the XMM-Newton process is not straightforward. 
The DDT is an especially  valuable tool, particularly when it can be employed at the outset of a mission to  demonstrate the scientific promise of joint observations.
This potential  is exemplified by the Planck and eROSITA follow-up observations. 
Nevertheless, the request for DDT necessitates the involvement of an experienced scientist from the other project.

Furthermore, joint programmes, particularly those initiated at the outset of a mission, present a valuable opportunity for collaboration. 
The NRAO-XMM-Newton joint programme utilising radio wavelengths, was established relatively late and appears to be utilised to a lesser extent than the NRAO-Chandra programme, which was established much earlier. 
Furthermore, the Fermi follow-up observations required a considerable period of time to gain momentum. 
One efficacious method for encouraging collaboration is to invite scientists specialising in the relevant field to participate in the review process.
This approach allows for a more profound comprehension of the panel members' perspectives and expectations, particularly with regard to the feasibility study. 
The communities of the gamma-ray telescopes: HESS and MAGIC serve  as illustrative examples of this phenomenon. In response to a considerable number of requests and publications, joint programmes were initiated.

\subsection{Evolution}

Figures \ref{EvoNum} and \ref{EvoTime} demonstrate a discernible progression in the expansion of the discovery space, occurring in three distinct phases, corresponding to approximately seven years duration for each.
The initial step was the allocation of longer observation times, which resulted in a significant improvement in the  S/N, with a factor of 10 increase. This was  intrinsic  to XMM-Newton and did not necessitate external collaboration.
The second phase of expansion was characterised by the coordination of observations and joint programmes. The possibility of undertaking such observations was prompted by the advent of new facilities external to the project, which provided the opportunity to extend the scope of the investigation. The requests for such programmes reflect the scientific potential and the possibility of addressing the same physical mechanism with two or more instruments, thereby demonstrating the value of such programmes in advancing scientific understanding.
The third category comprises TOOs. As with the preceding step, the potential for further discovery in this direction is contingent upon the environment, specifically the establishment of monitoring surveys for transients. 
The final two stages of expansion entail the incorporation of new communities, where initiating joint programmes early on is of the essence.

\section{Peer Review}
\label{PeerReview}

\subsection{Purpose and Functions of the Peer Review Process}

The primary role of the peer review process is to determine the observing programme for the upcoming cycle. For missions with long operational lifetimes, such as XMM-Newton, the review process also serves as the main channel of interaction between the observatory and its scientific community.

The XMM-Newton Peer Review must define the observing programme so as to maximise the potential for high-impact and transformative publications \citep{2023Natur.613..138P}. 
It is essential that the scientific community trusts the process and its fairness. Moreover, the selected observing programme must demonstrate its scientific merit to both the review members and the wider community. It must ensure a consistent stream of refereed publications and, critically, demonstrate value to funding bodies, thereby justifying mission extensions.

The review process also enables active involvement of the community in the observatory’s decision-making, thereby ensuring broad and sustained support for the mission. Such engagement strengthens ties between the mission and its user communities, including researchers from smaller countries and newly acceded ESA member states. For the vast majority of participating scientists, the XMM-Newton peer review is not perceived as an onerous duty, but as a constructive scientific event that is firmly embedded within the annual cycle.

A central principle is that outcomes must be free from bias, with gender, academic seniority, and native language being the most common sources of concern.

\subsection{Refining the peer review process over time}

XMM-Newton’s peer review procedures have been continuously refined, with the aim of minimising bias. Unlike other missions, which introduced reforms after decades of data collection, XMM-Newton has followed a bottom-up approach, optimising its procedures progressively since AO1.

Any changes were made only after extensive discussions with the chairpersons of the Time Allocation Committee (TAC), who oversaw four cycles each. Over the past 25 years, the XMM-Newton Peer Review has been chaired by distinguished scientists such as Malcolm Longair, Brian McBreen, Catherine Cesarsky, Marco Salvati, Peter Schneider, and Phil Charles. Moreover, feedback from panel chairs, Users’ Group meetings, and individual exchanges proved pivotal in shaping the process.

The initial XMM-Newton review followed the standard model, with a panel for each of seven scientific categories. The number of members per panel ranged from five to nine. Members rated proposals in advance, which were then discussed and decided upon collectively. Conflicts of interest required abstention from rating and discussion. Because of limited SOC staff, panels met on different dates, allowing the process to be managed with only ~1.5 staff members. Participants came from nearly all ESA member states.

The preliminary assessment was largely successful, though discussions highlighted areas for improvement. Some modifications, endorsed by the Science Working Group, were introduced. 
For AO2, the number of panels was raised to 15, thereby ensuring at least two panels for each scientific category. Each panel had three members. This helped manage conflicts of interest more effectively.

In AO3, the number of panel members rose to five, as panels of only three sometimes struggled, particularly when illness or travel issues arose. AO3 therefore adopted decentralised panel meetings, typically at each chair’s home institute. This model became standard for all subsequent AOs, with continuous optimisation based on challenges encountered during the XMM-Newton reviews and on experience gained through participation in reviews of other missions.

From AO3 onwards, the review adopted a mixed model: proposals are submitted by scientific category, with two or more panels per category. Members rate proposals beforehand; panels then meet independently to discuss and allocate time. Panels do not know the membership of others, as meetings occur at different dates and places. Finally, panel chairs meet to decide on large programmes, resolve duplicate targets, and adjust time allocation for joint facilities if necessary.

\subsection{Panel Composition and Conflicts of Interest}

The review is conducted by experienced astronomers, most of whom regularly publish in refereed journals. Many panel members also have direct experience with XMM-Newton data. While most hold PhDs, some do not, and senior figures—including faculty deans, institute directors, and journal editors—have participated as well. Panelists span all career stages, from graduate students to leading scientists, meaning that some reviewers may influence the careers of others and that some participants may be directly affected by their decisions. It is important to recognise that the review process involves individuals with a diverse range of attributes, some of which may be advantageous and others potentially challenging. Furthermore, it must be acknowledged that humans are social creatures, which can influence the dynamics of the review process.

Conflicts of interest are a persistent challenge. Panelists are highly aware of them and apply effective mitigation measures. However, some biases—particularly unconscious ones—are harder to detect. For this reason, XMM-Newton procedures were revised early to minimise their impact. Most of these biases can be attributed to concerns related to the non-disclosure agreement inherent in the refereeing process, namely that all discussions and materials within the review are kept strictly confidential. Such concerns may lead to biased ratings, reluctance to challenge a proposal, or difficulty in discerning the rationale behind a proposal’s arguments. Individuals respond differently to the same circumstances; consequently, factors that do not affect one panel member may influence the ratings of others. It is reasonable to suggest that personal unconscious biases may exert less influence than those introduced by the review process itself. The following examples illustrate how conflicts of interest and procedural issues may arise. 

Examples include

\begin{itemize}
    \item 
The panel member is the PI of a proposal that is currently under discussion.
In order to resolve the conflict of interest, it is standard practice for the PI to refrain from rating their own proposal and to absent themselves from the room during the discussion and decision-making process. In most  cases, the panel members with a conflict of interest adhere to the established norms and refrain from providing commentary on their own proposal.  
    Nevertheless, in numerous instances, proposals are accorded high rankings and accepted because that the PI's familiarity with the subject matter influences the assessments of the remaining panel members. 
    It is not possible to eliminate this bias within the context of the panel procedure.
    This may indicate a deep-seated social and cultural tendency to respect judges, which in turn raises the question of how judges should be evaluated.
    \item   A panel member who is a co-investigator (CoI) on a proposal that is currently under consideration. This conflict is resolved in the same way as the previous one, i.e., theh CoI refrains from rating their own proposal, and is absent during its discussion.
     In general, conflicts of interest -- and especially the implied bias -- are of lesser significance for CoIs than for PIs.
      In most cases, the proposal is regarded as an original research concept by the PI,  who tends to hold a strong opinion about it. 
      By contrast, the reasons for which a CoI  joins a proposal team are diverse and may, in some cases, include limited conviction in the proposal itself.
    \item 
    A member of the time allocation committee is the PI of a proposal that is evaluated by a panel other than their own. 
    In many peer-review contexts, it is assumed that such a PI does not introduce bias into the decision-making process, provided they are not part of the relevant panel. However, in several cases, it has become clear that their  involvement can influence the outcome. 
    It is reasonable to assume that this bias correlates with the PI’s scientific standing: in other words, a PI with a higher reputation and greater visibility is more likely to exert influence than one with a lower profile.
    \item 
     Two or more members of the panel hail from the same community. In certain instances, panel members may be reluctant to express opposition to a proposal 
with a PI from their community when another panel member from that community is present. 
In such instances, panel members may evince a lack of trust in the non-disclosure pledge pertaining to the panel discussion. It seems reasonable to posit that this bias correlates with the prominence of the PI within the community.
    \item 
    A member of a panel and the PI of a proposal are affiliated with the same organisation. 
     This can introduce a bias into the panel member's evaluation. 
For large research institutions, however, it is possible that the panel member is unaware of the PI's   current affiliation. Conversely, a group leader or institute director may have a significant vested interest in the success of a high-profile programme, such as a MYHP, with a PI from their institute.
    \item 
    Strategic conflicts of interest and biases:
The majority of panel members are also engaged as investigators in the call. It is therefore imperative that the procedure ensures that the decision reached by one panel does not influence the decision regarding  proposals discussed by another panel, in which  those same panel members are participating as investigators. 
  \item 
  Since  AO24, OTAC members are mandated to engage in a process of introspection regarding unconscious biases prior to the initiation of the evaluation process. This introspection is facilitated by 
a video on unconscious biases from the Royal Society\footnote{https://www.youtube.com/watch?v=dVp9Z5k0dEE}, along with a link to the analysis by \cite{2024arXiv240214075P}. The explicit objective of this introspection is to achieve gender parity in the success rate, that is, to ensure that the gender distribution of accepted proposals approximates that of
submitted proposals.
\end{itemize}

The XMM-Newton review is structured in a manner to ensure the presence of the requisite scientific expertise while simultaneously addressing potential conflicts of interest and biases between panel members and proposal PIs or CoIs, namely:
\begin{itemize}
\item 
Two or more panels are established for each scientific category, thereby enabling the mitigation of significant conflicts of interest (e.g., those between PIs and CoIs).
  \item 
  Each panel comprises five members whose home institutes are based in different countries.
This composition helps mitigate potential biases arising from shared national or institutional  affiliations.
    \item 
    Each panel 
includes a core of expertise in X-ray astrophysics, as evidenced by 
first-authored articles on X-ray spectra. This ensures the requisite technical expertise.
\item 
Panel members are not informed of the identities of the members of the other panels. 
\item 
The panels convene on different dates and in different locations, thus reducing the potential for bias arising from knowledge of other panel members.
\item 
The pre-defined and fixed budget of observing time for each panel 
helps mitigate potential biases 
stemming from the individual research interests of panel members.
\item
The objective is to ensure participation by 
at least one review member from each ESA member state, thereby facilitating connections with all communities. 
\item 
The majority of panels are
composed of investigators of previous AOs, thereby enabling the actively proposing community to determine the fate of their proposals. 
\item 
To ensure the inclusion of a sufficient number of high-quality proposals and to allow for the coverage of smaller areas, approximately 40 proposals are 
assigned to each panel.
\end{itemize}

In light of the aforementioned configuration, the XMM-Newton review  identifies the most meritorious proposals. 
Each submission is evaluated fairly and comprehensively with due consideration given to their entirety. In the event of a low number of submissions, no penalties 
are applied.
The panel's technical expertise enables the identification of potential issues 
regarding feasibility in the proposals submitted.

\subsection{Panel Dynamics and Scientific categories}

The XMM-Newton AOs have demonstrated a sustained oversubscription rate of over 5.5, with an average of around 6.9. 
The distribution of available observing time is designed so that each panel 
faces the same oversubscription factor.
In addition, each panel in the same scientific category evaluates the same number of proposals, ensuring that all proposals are reviewed under identical constraints of proposal count and observing time. 
As the average requested observing time differs between the different scientific categories,  only the oversubscription in terms of time can be kept constant across all panels.
The number of proposals is sufficient to ensure that each panel is presented with a pool of high-quality submissions, 
thereby ensuring a similar level of competition for all proposals. 
Moreover, the distribution of 40 proposals ensures that smaller subareas 
receive adequate consideration.  
Most panels divide proposals into subgroups, clustering those that address related questions.  
Subsequently, each proposal within the subgroup is assigned a priority level, which is a relatively straightforward process.  
The final programme is defined on the basis of the priorities achieved in the subgroups. This is a more challenging process, as it requires
balancing disparate scientific questions or object classes. 
In the event that a single proposal encompasses a specific area, there is a risk of it being under-represented in this process. It is therefore essential to ensure that there is a sufficient number of proposals, even in 
relatively narrow research areas.

At the time of writing this paper, proposals are evaluated in accordance with five scientific categories.
Over time, the number of categories has been reduced from seven to five reflecting a notable decline in 
submissions in certain areas of research. 
To illustrate,
interest in the study of galaxies declined 
following the observation of nearby galaxies by XMM-Newton. Owing to its superior
 spatial resolution, Chandra is the optimal instrument for observing more distant galaxies 
for population studies.
The reduction in categories, which involves a regrouping of topics, allows XMM-Newton to maintain its policy of 
assigning at least two panels to each scientific category. 

The targeted scientific categories ensure that 
most panel members are experts in the relevant field, thereby keeping the discussion focused on scientific aspects. 
It is important to recognise that a proposal is fundamentally distinct from an examination or a test in a university course. Consequently, although all statements 
in a proposal may be accurate,
a panel may elect to assign a higher rating to another proposal and to allocate observing time to it. 
The scientific value of a research proposal must be 
assessed in the context of the panel members' expectations regarding future developments in the field.  It is not uncommon for scientists to hold 
differing  views on this matter. 
It can be argued that the highest-rated proposals may be
regarded as speculation on the future, offering a high a high degree of potential gain but also a corresponding level of risk.
The panel members with expertise in 
XMM-Newton data analysis 
are responsible for identifying feasibility problems and determining whether the archive 
already contains observations supporting the scientific case. 
In many cases, overly optimistic expectations of the signal-to-noise ratio constitute a substantial impediment, particularly for extended sources. 
It is therefore crucial that the panel includes members with the necessary technical expertise, as the SOC, despite having the necessary expertise, cannot provide detailed technical evaluations—such as feasibility studies—due to limited personnel resources 
Moreover, for 
the same reason, the SOC cannot ensure adequate representation across all areas of expertise.

.  

\subsection{Selection of Panel members}

The majority of OTAC members are selected from the PIs and CoIs of recent calls for observing time, following cross-checking with the
Astrophysics Data System (ADS) publication list. 
Consequently, the proposing community is responsible for determining the observing programme.
It would be scientifically inadvisable for the OTAC to remain exclusively within 
a single scientific community. 
Therefore, scientists working with data from other wavelength ranges are consistently invited to participate. 
It is essential  that such scientists have
overlap in physics, instruments, or analysis methods. 
Experience has demonstrated the value of including scientists with backgrounds in radio, gamma-ray, or TeV energies, as well as senior colleagues from ESA advisory bodies. Astronomers whose research interests lie in other physical areas and who employ disparate analysis methods may agree to serve on a panel once, but frequently decline 
a second invitation. 
OTAC members are invited to provide assistance on two consecutive 
calls, after which they will be replaced. 
It should be noted that generally the meetings are not held at an ESA facility and that SOC staff are not involved in their organisation.
A sophisticated software tool is provided to facilitate both the rating process and the panel meeting.
It is therefore beneficial for some panel members to have prior experience of the process and tools in question.
In a formal request, the Science Working Group has requested that panel members be rotated.

It is intended that at least one panel member be drawn from each ESA member state.
Nevertheless, this is not always 
viable for smaller countries lacking the necessary X-ray, astronomical or physical research communities.
It is the aim to select  the panel chairpersons in accordance with the principles of gender parity.
Moreover, the objective is to ensure that a minimum of 30\% of the panel members are female, in accordance with the minimum threshold required to guarantee balanced discussions. 
As demonstrated by \cite{2024arXiv240214075P} Figure 10 , only 20\% of proposals are submitted by senior female researchers. 
This presents a significant challenge in identifying a sufficient number of females as panel members, let alone panel chairpersons, given the requirement for higher seniority. This finding is particularly salient in the context of the high demand for female astrophysicists on various committees. This phenomenon may 
indicate a state of overexploitation, potentially motivating experienced female astrophysicists to decline invitations to become chairpersons.

Italy has the highest number of active X-ray astronomers in Europe, by a significant margin.
Given the necessity for expertise in X-ray analysis and publication, in conjunction with the restriction of one panel member per community, in the majority of countries, a substantial number of active X-ray astronomers were invited to participate 
more than twice throughout the mission's duration. This phenomenon was less evident among Italian astronomers, reflecting the size of the Italian community.

\subsection{Evaluation process}

The evaluation process is designed to ensure that panel members are free from any conflicts of interest or bias regarding the PI.  
Within panels, conflicts of interest and bias concerning CoIs are also eliminated. 
In most cases, this can be achieved by utilising the two or more panels available for each scientific category. 
When it is  necessary to transfer proposals between scientific categories, we aim to move more than one proposal in order to avoid  bias against small numbers of proposals.  This has been remarkably successful.
In the context of large programmes, it is inevitable that a very small number of biases 
concerning co-investigators may arise at the chairpersons' meeting. 
The resolution of institutional, strategic, and personal conflicts of interest is achieved through the 
use of classical methods. 
In such instances, the remaining potential for bias is relatively minimal and can be 
considered negligible. 
The open naming of the proposers enables panel members to proactively identify any potential biases that were previously unknown to the SOC during the proposal allocation process.  
The public disclosure of investigators' identities ensures that all potential conflicts of interest and residual biases are transparent to panel members, thus facilitating their 
detection and management. 
It is worth noting that, in contrast to NASA missions, the ESA does not provide financial support for data analysis and publication for successful investigators.
It is therefore requested that investigators provide commentary on previously awarded proposals that have resulted in observations.
This approach allows the avoidance of further investment of observation time in the event of a substantial volume of unpublished data.

A large programme is defined as a research project 
requiring at least 300 ks of observing time and is identified as such by the PI. 
Evaluation is conducted by two panels, which may forward selected large programmes to the chairpersons for consideration.  
Chairpersons may not be PIs of large programmes.
Up to two chairpersons may be drawn  
from institutes of major ESA member states or the USA, 
from different scientific fields 
helping to reduce potential community bias. 
The panel chairpersons  meet 
in a final session, chaired by the OTAC chairperson, to deliberate and reach a decision on the large programmes.

Very Large Programmes were offered from AO7 to AO16 and their review process 
followed the same procedures used for Large Programmes.   Multi-Year-Heritage Programmes (MYHP) have been 
offered in AO17, AO21 and AO23 (Sect.~\ref{Large}).
MYHP proposals are evaluated by a dedicated panel, the Senior Review Panel, composed of highly experienced scientists, most of whom have previously served as panel chairpersons.
To ensure optimal panel composition and the necessary expertise, the community is invited to submit a letter of intent approximately six months before the proposal deadline.

It is important to note that the evaluation criteria for standard and Large Programmes differ significantly from those applied to MYHP. 
For standard and Large Programmes, a hypothesis is formulated and the required statistics determine the exposure time; the requested exposure time often represents a central point of discussion.
By contrast, MYHP evaluations typically do not involve hypotheses\textbf{.} The scientific strategy for MYHP is to cover the essential portions of the observational discovery space, ensuring a broad and systematic approach. 
As discussed in Section~\ref{Large}, low priority time is often scientifically meaningless for Large or MYHP programmes, while the available high-priority time is distributed among Large, Very Large, MYHP and guest observer programmes in a manner that ensures each receives the same level of oversubscription. 
This allocation has resulted in approximately 37\%, 16\%, and 45\% of high priority time being assigned to each group, respectively.

\subsection{Outcomes and Reflections on Peer Review Models }

An analysis of past review outcomes demonstrates that the XMM-Newton process has consistently produced results with low levels of bias—comparable to those achieved by other observatories only after adopting double-anonymous reviews. This section examines different review models and highlights the advantages of XMM-Newton’s single-anonymous, panel-based approach, particularly for early-career researchers. It also considers how maintaining visibility can foster recognition and engagement within the broader scientific community.

Recent studies have examined the results of time allocation peer reviews at several astronomical facilities, including the European Southern Observatory (ESO) \citep{2016Msngr.165....2P}, the Hubble Space Telescope (HST) \citep{2014PASP..126..923R}, the Atacama Large Millimeter/Submillimeter Array (ALMA) \citep{2020PASP..132b4503C}, the Canadian Time Allocation Committee \citep{2018SPIE10704E..0LS}, and the National Radio Astronomy Observatory (NRAO) \citep{2016arXiv161104795L}. 
These studies have revealed significant gender and academic age biases, with differences in success rates reaching up to 39\% \citep{2016Msngr.165....2P}. In response, 
several facilities have implemented double-anonymous refereeing processes to mitigate these effects.

\cite{2024arXiv240214075P} analysed 20 cycles of single-anonymous peer review for XMM-Newton (AO2–AO21), covering two decades of data. Their findings show that the process yielded outcomes with low bias, comparable to those achieved by other major facilities only after switching to double-anonymous review in recent years.  
Some Announcements of Opportunity (AOs)
showed higher success rates for women than men
while in others the reverse was true,
indicating that gender-related success rates fluctuate from cycle to cycle. This level of variability is comparable to that observed in HST outcomes in the years following the implementation of double-anonymous procedures. The resulting gender ratios (1.05 to 1.15) are consistent with those reported for HST after the transition.

The XMM-Newton review process also shows a strong success rate for early-career researchers: approximately 37\% of submitted proposals by individuals who had not yet obtained a PhD were accepted \citep{2024arXiv240214075P}. This is comparable to the HST results for first-time proposers under double-anonymous review. Moreover, no significant bias was observed in the language used \citep{2023Natur.619..678L}. Country-specific figures \citep{2024arXiv240214075P}—for example, the United Kingdom (43.1\%), Italy (44.3\%), Germany (42.5\%), and France (48.0\%)—further support the consistency of the review outcomes. 

The analysis by \cite{2024arXiv240214075P}, was complemented by the SOC analysis using data from more recent years. Together, these combined analyses demonstrate that the XMM-Newton review has consistently yielded homogeneous results over the past 25 years, which are comparable to the outcomes attained by other major facilities only in the recent years following the implementation of the double-anonymous review.

An additional consideration regarding the single- versus  double-anonymous system is the size and structure of the X-ray astronomy community. This community is relatively modest in size.  According to estimates by \cite{Ness2}, the number of active lead researchers (first authors) in the field is around 570. Given the diversity of research domains, it is common to find only three or four research groups 
focusing on a given topic. As a result, experienced researchers can often identify the PI or research group behind a given proposal, making a truly double-anonymous review difficult to achieve in practice.

Moreover, scientific ideas are highly sensitive, especially novel and potentially transformative ones. These ideas often face rejection upon first submission, sometimes due to target selection or feasibility issues that require further development and resubmission. However, even when proposals are rejected, other researchers tend to associate the idea with the PI, who is recognised as the intellectual originator. This implicit attribution is almost automatic and difficult to avoid: once the review process has 
concluded, human memory cannot be erased -as a computer hard 
drive is-. All parties involved in the discourse associate the concept with the name of the PI, thereby ensuring that the 'priority' is upheld. Invited talks at XMM-Newton workshops or conferences are frequently offered to individuals who propose innovative ideas, further strengthening the link between the idea and the scientist. 
In other words, regardless of the outcome of the review process, research ideas tend to remain associated with the PI. For early-career scientists, this visibility is crucial in gaining recognition and advancing their careers. 
In a fully double-anonymous system, it is worth questioning whether the same level of recognition and priority attribution would be preserved.

It is noteworthy that no significant language or gender biases have been observed in the XMM-Newton review process, and early-career researchers demonstrate notably high success rates \citep{2024arXiv240214075P}. These findings suggest that panels  
evaluate proposals primarily on their scientific merit, regardless of the proposer's background or career stage. Moreover, the single-anonymous process allows reviewers to be aware of a PI’s linguistic or professional context, which can actually help them interpret the intent of a proposal more accurately—particularly when  the language lacks precision, a fundamental aspect of scientific discourse—thereby mitigating potential bias against non-native English speakers. Similarly, in several cases, arguments in favour of a proposal have explicitly acknowledged the PI’s early career stage, recognising the value of encouraging promising new contributors to the field.

Nevertheless, both the XMM-Newton review system and the double-anonymous approach used by HST, despite efforts to eliminate bias, still show gender-related differences, with male success rates 5 to 15 percent higher than those of female applicants (\cite{2024arXiv240214075P}). A statistical assessment of factors external to peer review—based on individual cases—could be valuable. One hypothesis 
to test is whether male researchers are more likely than female researchers to view the proposal process as a competitive endeavour, with the explicit aim of winning. Such a perception could influence the strategies adopted and, ultimately, affect the outcome.

There is a recent shift toward distributed peer review, where PIs evaluate a few other proposals. While this approach is more cost-effective,
it reduces the professional recognition and value of being a reviewer and may discourage participation -- especially as it adds work without clear acknowledgment, and success rates remain low. In contrast, XMM-Newton’s 
panel-based review system is not particularly resource-heavy and is seen as a mark of distinction, especially for early-career scientists. Furthermore, the panel-based review 
facilitates meaningful interaction both with the SOC and among researchers, reinforcing transparency, collaboration, and community engagement.

\section{Education and Early Career Scientists}

\subsection{Indicators of Impact on Education and Early Research}

XMM-Newton plays a significant role in the education of astrophysicists. 
This is evidenced by the 540 PhD theses documented by the SOC (https://www.cosmos.esa.int/web/xmm-newton/phd-theses), the vast majority of which are based on XMM-Newton data, with the remainder focusing on technical aspects, such as calibration, hardware, or software development. 
It is important to note that this figure represents a lower limit, as it includes only those PhD theses that have been reported to the SOC — typically by the author or supervisor. 
As is generally the case, it is assumed that the number of lower-level academic degrees, such as Bachelor's or Master's theses, is significantly higher than that of PhDs, although there is no record to quantify this.

The role of XMM-Newton in education and the early career stages is further supported by the activity period of authors of papers in refereed astronomical literature based on XMM-Newton data (see Ness et al., 2003) (Fig. 6). Around three-quarters of the first authors publish for six years or less. 
These temporal patterns are consistent with conventional career trajectories in scientific research, which typically involve a three- to five-year doctoral qualification, followed by one or two brief periods of one to three years of postdoctoral research. 
Another indicator of the impact on education is the number of proposals submitted by researchers at the start of their academic careers. Here, "academic age" is defined as the number of years since obtaining a PhD, or as zero or negative for those without a PhD.
The distribution of PIs according to academic age also reflects this trend, as illustrated in Fig. 9 of \cite{2024arXiv240214075P}.  The mean academic age of PIs is 10.9 years, with the highest number of proposals submitted two to seven years after PhD completion.  

XMM-Newton observations are conducted between six months and one-and-a-half years after a proposal has been submitted.
This timeframe is too long for a PhD thesis based on data requested by the candidate.
Most PhDs are based on data proposed by supervisors or made public in the archive.

\subsection{Organisation of the X-ray Community and Career Prospects}

The substantial volume of research activity carried out by students and early-career scientists is indicative of the organisational structure of the X-ray astronomy community. This community is characterised by a lack of large institutes dedicated to this field of research. 
The same is likely true of astrophysics more broadly, where commercial and industrial interests and investments are negligible or non-existent. Consequently, early-career scientists face intense competition when trying to secure academic positions.

In this context, peer-reviewed publications are undoubtedly the primary criterion for the recruitment of scientific and academic staff, with preference given to those with significant and broad impact, as evidenced by citations. 
Thanks to the long duration of the XMM-Newton mission and its continuous user support, the XMM-Newton Archive has grown into a valuable and extensive resource for scientific publications. 
 Even well-known sources possess data that remains unpublished due to six instruments observing concurrently. Nevertheless, despite the significant potential of the public data available from the archive, new, pristine data are more likely to result in novel and unexpected discoveries.

 Formulating hypotheses and testing them with new, unobserved data is a vital step in the scientific development of early-career scientists. 
XMM-Newton observing time proposals are one way of acquiring novel observational data. 
Unlike the publication of articles, proposals involve convincing a panel of scientists of the value of novel scientific ideas. 
In the United States, participation in the proposal process is highly regarded as a significant achievement, with a large proportion of funding contingent on the success of these proposals. 
In Europe, scientists who have successfully led large research programmes are well placed to secure academic positions. 
PhD researchers are often well placed to formulate proposals, thanks to their in-depth knowledge of a specific field and the insights they have gained through their research. 
However, this advantageous position can be misleading, as the choice of PhD topic is often made by the supervisor, with an eye on the available data. Subsequent data in the area may therefore share the limitations discussed in the section on the discovery space.  
Therefore, it is crucial to use the postdoctoral phase to expand methodological expertise and broaden the scope of scientific research. 
Academic institutes generally hire people based on long-term scientific potential, rather than specific skills and knowledge.

A large number of XMM-Newton publications and observing time proposals by doctoral researchers and early-career scientists can enhance their chances of continuing in academia and sustaining active research groups. This, in turn, contributes to a healthy and dynamic research environment.  

\section{Scientific User Support}

For a long time, X-ray astronomy was regarded as a distinct and challenging field, the domain of a restricted group of astronomers with specialised expertise.
The main reason for this was the poor energy resolution of previous X-ray detectors, which prevented the direct analysis of spectral features such as emission lines in the measured spectra.
To overcome this limitation, X-ray astronomers use detector responses and employ forward folding and statistical comparison to analyse the resulting spectra.
To achieve this, the entire spectrum, including the continuum, must be incorporated into the models. 
This methodology differs markedly from that used to analyse classical optical spectra. 

As a large astronomical observatory, XMM-Newton provides extensive scientific support.
The fundamental principle is that any scientist, regardless of their background, should be able to propose, analyse, and publish observations made with XMM-Newton.
An analysis of all peer-reviewed articles using XMM-Newton data \citep{Ness2} revealed that 51\% of first authors published only during one year, with most of these authors publishing only one XMM-Newton article.
This high proportion of one-time authors illustrates the successful implementation of this concept, making XMM-Newton observations an optimal choice for early career researchers — including PhD students — as well as for broader educational initiatives.

The XMM-Newton SOC is the primary point of contact for the scientific community, providing support in various areas.
The XMM-Newton Helpdesk provides timely assistance for a wide range of individual queries, including technical issues, observation planning, data analysis and scientific interpretation.
To facilitate the efficient and effective use of the mission's capabilities, the SOC maintains and updates extensive documentation covering mission operations, instrument characteristics, calibration procedures, and data analysis methods. 
A suite of dedicated software packages, tools, and online resources is also provided to support researchers at every stage, from proposal preparation to data exploitation.
 The SOC manages the processing of observational data, from raw telemetry to science products, performs quality control and ensures the long-term preservation and accessibility of the mission’s data products. Regular instrument calibration and performance monitoring are carried out to maintain the scientific quality of the mission’s output (see the next section).
 
 The proposal system only requires a few source parameters (coordinates, anticipated X-ray flux, fundamental spectral model and optical magnitude) to define and validate the observational modes for the X-ray instruments, as well as to adjust exposure times and add mode-dependent overheads.
 This enables scientists without expertise in X-ray astronomy to swiftly determine the necessary configuration.
For each observation, regardless of the priority assigned by the review panels, the SOC verifies the selected instrument configuration to ensure safety and optimise scientific performance.
The PI is then contacted and informed of any mandatory changes, as well as recommendations for scientific optimisation. 
Once the PI has confirmed these changes, the final configuration is established and the observation is incorporated into the planning.  
Any scientific time constraints are taken into account during the scheduling process.
All data obtained from the observations is processed through a standardised pipeline.

 The SOC, in collaboration with the Instrument PI teams, assumes responsibility for the calibration of instruments. 
 In partnership with the SSC, the team develops and provides the Science  Analysis System, SAS,  which enables customised  data analysis.
 The SSC produces catalogues of all X-ray sources detected with XMM-Newton \citep{2003AN....324...89W, 2009A&A...493..339W, 2016A&A...590A...1R, 2020A&A...641A.136W, 2019A&A...624A..77T, 2020A&A...641A.137T, 2008A&A...480..611S}. 
 Furthermore, the OM instrument team, coordinated with the SOC and SSC, releases catalogues of optical and ultraviolet sources \citep{2012MNRAS.426..903P, 2017MNRAS.466.1061P}. 
 The NASA/Goddard Space Flight Center (GSFC) XMM-Newton Guest Observer Facility (GOF) published a another OM catalogue \citep{2008PASP..120..740K} in 2008
 These resources are becoming increasingly utilised, as evidenced by the growth in its usage, as illustrated in Fig.~\ref{fig2} (class 5 in red). 

The XMM-Newton Science Archive (XSA) fulfils the following primary functions: firstly, the distribution of proprietary observing data; secondly, the provision of access to public data; thirdly, the provision of an overview of the scientific content of selected data; fourthly, the support of work for the Science Operation Centre; fifthly, the access to  the primary catalogues; and sixthly, the provision of upper limits. 
The XSA has been optimised for use by scientists with no specific knowledge of XMM-Newton. 
While optimised for use by such scientists, it also offers different modes for expert users. 
In astronomical research, the most frequently used search criteria are the target name or celestial coordinates. The XSA also allows users to specify additional parameters, such as the observation identifier or the instrument configuration. The archive provides access to individual pointed observations, pipeline products, slew data, catalogues, and upper limits. Retrieved entries allow for an initial evaluation of images, spectra, and light curves, supporting an assessment of their suitability for further scientific analysis.

From 2004 to 2013 (XSA versions 2.5 to 7), the archive offered a range of interactive, online analysis tools. These included reprocessing data, filtering processed files, and extracting spectra and light curves, together with the necessary auxiliary files. All of this was based on the latest SAS and calibration data. This functionality was suspended between 2013 and 2017, following the migration of the XSA to a fully web-based interface. Since 2017, the archive has offered this capability again, through the Remote Interface for Science Analysis (RISA). These developments can be seen as precursors to today's cloud-based science platforms.
A review of publications up to 2020 reveals that 28\% of articles use non-archival data, 20\% use partly archival data, and 50\% use archival data \citep{2024arXiv240212818D}.

Active researchers working with XMM-Newton data were supported through an electronic newsletter, the "XMM-Newton News", which replicated the format of the NASA Great Observatories newsletters. The purpose of the newsletter was to ensure that researchers were informed of all changes that were relevant to their scientific work in a timely fashion. 
The scope of these updates encompassed a wide range of developments, including updates to calibrations, new versions of analysis software, new source catalogues, planning for announcements, conferences and workshops, announcements of letters of intent, opportunities and XMM-Newton organised conferences and workshops, results of calls for proposals and OTAC decisions, calls for input for the XMM-Newton PhD statistics and prizes, and pointers to conferences that explicitly mentioned XMM-Newton in their subject matter. The newsletter was accompanied by the NASA GOF published GSFC XMM-Newton GOF Status Reports, which included equivalent content, in addition to US-specific information, such as funding for US-based proposers. 

Providing timely information is highly beneficial to scientific research, particularly in cases involving large samples that require months of analysis. For example, knowledge of a calibration update or an input source catalogue update may require the work to be redefined or frozen at the previous status, which could lead to a subsequent request for an analysis update from a referee during the publication process. Failure to receive such updates in a timely manner can result in weeks of work having to be redone. Following the discontinuation of XMM-Newton News, all relevant information can be found on the XMM-Newton website, where the SOC aims to highlight changes on the homepage. Announcements of opportunity and conferences are published in ESA Science News. 
XMM-Newton GOF status reports continue to provide some of the content that was previously available in XMM-Newton News. 
The NewAthena Community Newsletter (https://www.the-athena-x-ray-observatory.eu/en) has announced updates to the source catalogues and software. It is reasonable to assume that there is considerable overlap between scientists actively working with XMM-Newton data and the NewAthena community. However, it is also reasonable to assume that there are active researchers who are not part of the NewAthena community, given that the anticipated launch date falls beyond their career horizon.

The handling of proposals and the planning of XMM-Newton observations has become increasingly complex over the years, particularly with the introduction of joint programmes, many of which require simultaneous observations. The number of target-of-opportunity and complex monitoring programmes has also grown significantly.
To address these challenges, the SOC has developed internal tools to automate many tasks, including those introduced specifically to meet new scientific requirements from the community. For instance, gravitational wave alerts are now processed automatically, enabling the SOC on-call scientist and the Project Scientist to be notified rapidly when a potentially interesting event occurs.
Despite these automation efforts, the SOC continues to contact each PI individually. In collaboration with the SSC, the SOC also performs visual inspections and produces quality reports for all processed datasets. 
These personalised efforts are highly valued by the scientific community.

\section{Calibration}

As XMM-Newton observes simultaneously with five distinct X-ray instruments, the maximum amount of information can only be obtained by considering all five instruments together. 
Since 2024, the calibration of these instruments has been such that a combined analysis is the recommended optimal approach.
This has been achieved by aligning the effective areas of all the X-ray instruments with those of the pn instrument.
The pn was chosen as the reference instrument because its effective area has remained consistent over time. No a priori preference was given to the 'pn' calibration over the others. 
Individual instrument calibration is still maintained and can be accessed through the selection of specific parameters during the processing stage.
Patching was conducted because the individual instrument-specific calibrations did not converge to a single calibration despite an extensive investigation period. 
Regarding cross-calibrations with other X-ray satellites, that with NuSTAR is particularly significant given the extensive joint and simultaneous observations conducted by NuSTAR and XMM-Newton.

The calibration process primarily aims to support users in interpreting their instrumental data as effectively as possible.
 This involves ensuring that the SOC, in collaboration with the instrument teams, provides publicly accessible calibration. The SAS enables users to retrieve the calibration files and apply them to the data.
In most cases, researchers can analyse XMM-Newton data. 
without undertaking a calibration process themselves.
Around five per cent of the total observing time  is dedicated to calibration observations. 
However, it is unclear whether scientists will ever be fully satisfied with the calibration for missions such as XMM-Newton. 
As well as constraints on human resources and limitations of the available calibration data, there are also intrinsic limitations of the instruments themselves and frequently overlooked fundamental principles.

\begin{figure}[t]
\centerline{\includegraphics[angle=0,scale=1.7]{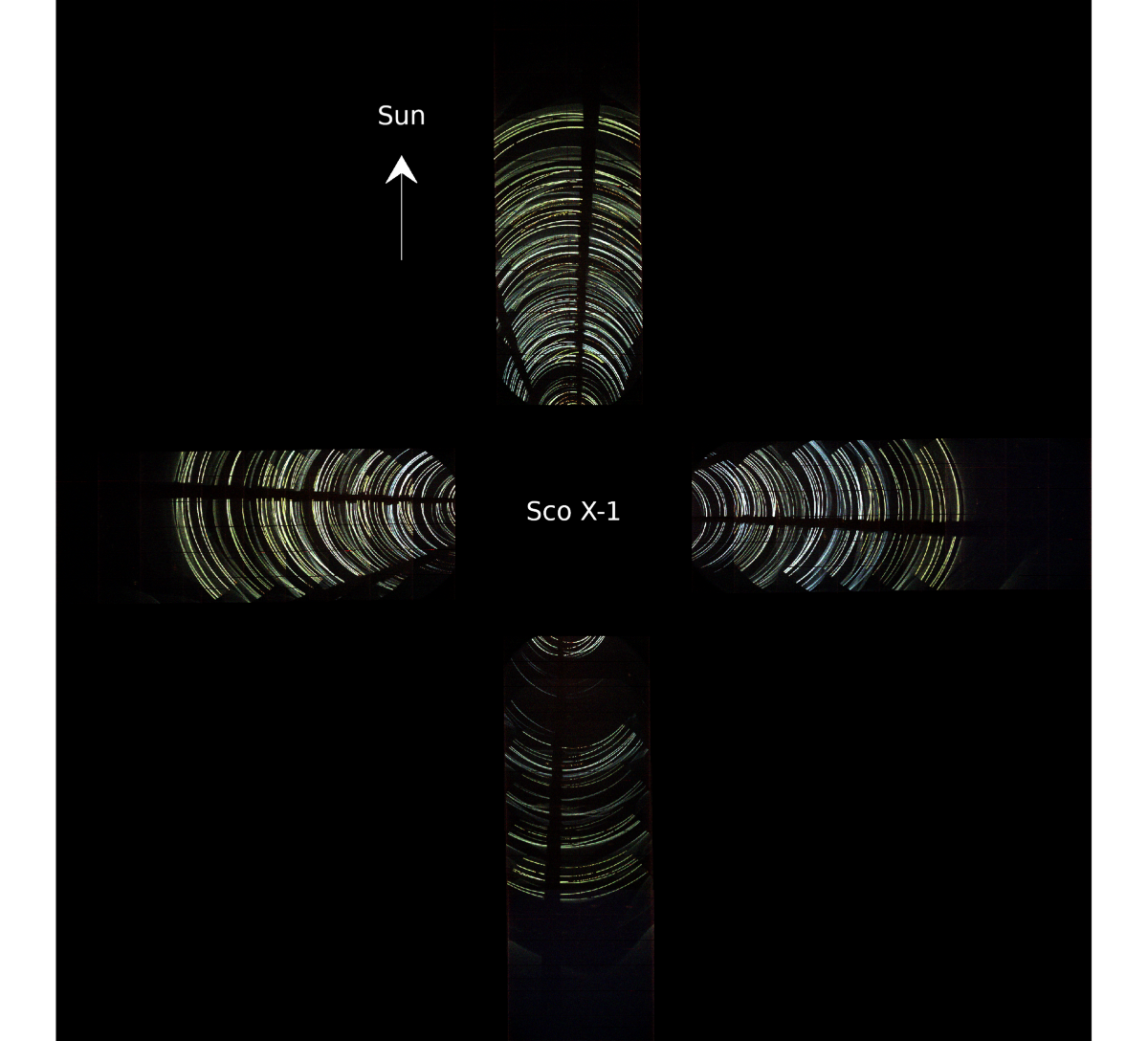}}
\caption{Off-axis observations of SCO-X1 to characterise mirror deformations\label{Sco}.}
\end{figure}

Calibrating the XMM-Newton X-ray instruments presented several challenges and limitations.
The following points are worthy of note:
\begin{itemize}
\item
      XMM-Newton had limited ground calibration due to various constraints, e.g. time and the lack of a sufficiently large long-beam facility to accommodate the complete mirror assembly.
\item 
     The absence of a celestial standard candle at X-ray wavelengths considerably impacts both the absolute effective area calibration and the shape of the effective area.
    A discrepancy of up to 38\% in the effective area at higher X-ray energies, when comparing XMM-Newton, eROSITA and Chandra based on observations of galaxy clusters, illustrates the ongoing challenges in this field. Further details can be found in the paper by \cite{2024A&A...688A.107M}. 
    \item  
     Due to the significant improvement in detector functionality, many of the celestial sources previously used for calibration now require a revised scientific interpretation. Consequently, the calibration process could not be separated from the astrophysical investigations specific to each source. 
    \item 
It is imperative that calibration requirements encompass a predefined verification procedure.
       Although specific astrophysical sources can be used for time and energy calibration, the lack of standard calibration sources for effective area poses a significant challenge in X-ray astronomy.   
    \item 
      The quantity of statistics that can be accumulated in calibration observations after launch is significantly limited because all XMM-Newton X-ray detectors are susceptible to photon pile-up.
    Photon pile-up occurs when multiple events deposit their energy in a single pixel prior to readout.
      Therefore, it is evident that utilising bright sources is not a viable approach for acquiring high statistics within a limited time frame. 
     For example, sources observed in pn full-frame mode must have a total flux of less than approximately three photons per second.
    In particular, calibrating the higher energy parts of the spectra presents a significant challenge, given that both the effective area and the celestial source flux are low in this range. While the full energy range contributes to the pile-up effect, only a small number of photons are collected within the energy range of interest and these are susceptible to contamination from piled-up events.
    Similarly, the effective area for off-axis positions is characterised by a reduction in the effective area and an extended, non-circular point spread function shape. 
      \item 
      Due to the extended operational lifespan of XMM-Newton, the onboard radioactive calibration sources have diminished in intensity due to radioactive decay. Consequently, the calibration process is gradually shifting towards utilising background fluorescence lines generated within the camera or observing celestial sources. 
     \item 
       Given the 25-year duration of the XMM-Newton mission, it is clear that classical constant calibration sources, such as supernova remnants which are typically  around 400 years old, cannot be considered constant.
    \item
Each XMM-Newton mirror module is composed of 58 mirror shells.
    It is not feasible to ascertain the existence of, or calibrate the potential deformation or direction of, the optical axis of each mirror shell in orbit (see Figure~\ref{Sco}).
    As each mirror shell reflects light at a different angle, the effective area of each shell exhibits distinct energy dependence.
    Any divergence from the pre-launch configuration of the mirror shells in terms of performance or position will result in inaccuracies in the effective area calibration.     
    \end{itemize}

The detectors in the EPIC and RGS cameras are charge-coupled devices (CCDs). 
Their response characteristics are constantly changing due to environmental factors such as particle incidence and accumulated photon flux. 
These changes include CCD charge transfer inefficiency (CTI), as well as gain and redistribution in the case of MOS cameras.
These changes must be corrected using time-dependent energy calibration. 
Furthermore, data collected during the mission revealed additional instrumental effects, such as gain dependency on particle rates for pn and CTI dependency on photon rates for pn fast readout modes.
While the weakening of the on-board radioactive calibration sources may be partially offset by celestial sources, compromises nevertheless need to be made in the energy calibration.

CCD CTI and gain are typically described in terms of CCD readout columns and rows. However, for charge traps, it is preferable to describe them at the per-pixel level.
Due to the limited and declining quantity of available calibration data, a more comprehensive characterisation of these effects is necessary over larger detector areas, up to the level of the CCD or chip. This approach will inevitably result in residual spatial inaccuracies in the energy scale calibration.

Furthermore, using windowed readout modes due to the restricted area being assessed, as well as the additional effects of illumination on non-telemetered detector regions, results in insufficient data to enable accurate, dedicated, mode-dependent calibration. In these modes, certain components of the energy calibration are derived from the corresponding instrumental full-frame modes.
 
The limited calibration data leads to increasing degeneracy between the CTI and gain effects, making their separation more approximate and resulting in greater inaccuracies in the energy scale calibration. This has implications for both MOS and pn, particularly with regard to the in-orbit calibration of particle background-dependent gain.

Although originally developed for high-precision timing, the pn Timing and Burst mode are also used to characterise the spectra of the brightest sources.
The high photon flux rates significantly impact CTI, necessitating dedicated, rate-dependent calibrations. However, accurately estimating the actual incident photon flux remains challenging due to limitations in the telemetry bandwidth, resulting in information being lost below the instrumental threshold. Furthermore, the variation of flux in the physical columns across the point spread function has not been considered due to insufficient calibration data. Additionally, combining physical pixel charges into macro pixels in the pn Timing mode introduces inaccuracies in event energy accuracy.
The net effect of these factors is an inaccuracy in the Fast mode energy calibration.

In X-ray astronomy, particularly with regard to prospective X-ray missions, adherence to the following guidelines would be advantageous:    
\begin{itemize}
\item 
It is essential to establish a consistent standard candle for calibrating the energy flux at specific energies, for example a spatially compact cluster of galaxies,. 
\item 
When designing the instrument, it may be advisable to consider background fluorescence lines for energy calibration, with the aim of minimising interference with astronomical measurements. 
\item 
 It is not possible to calibrate a mission involving a large number of mirror shells or segments based on in-situ or in-orbit observations alone. 
 In the absence of a celestial X-ray standard, it would be sensible to incorporate a calibration support mission that allows comparison with a calibration source, as part of the mission plan. 
\end{itemize}

\section{Workshops and Conferences}\label{conf}

The XMM-Newton SOC organises a major triennial conference called 'The X-ray Universe', aiming to hold the meeting in the capitals of ESA member states.
These conferences cover a wide range of high-energy astrophysical topics. 
The morning session comprises two plenary sessions, while the afternoon session features four parallel sessions, each addressing a different topic. 
In the years between major conferences, the SOC organises workshops at ESAC focusing on a specific high-energy subject comprising solely plenary sessions.

Given ESA's status as an intergovernmental organisation comprising 18 member states, we have ensured that members of the scientific organising committees and invited speakers are affiliated with a diverse range of countries. 
It is recognised that this is often more challenging to achieve for a topical workshop than for a conference such as X-Ray Universe.

The selection of invited speakers plays a pivotal role in defining the identity and appeal of a conference and thus influences scientists' decisions to attend. 
The process of identifying invited speakers has evolved considerably over time. 
The most practical approach is to ask the members of the Scientific Organising Committee for suggestions on an individual basis.
Asking committee members individually allows the number of suggestions for a potential speaker or topic to be used as an indication of the likelihood of an engaging presentation.  
Based on this list, the Chairperson, supported by the PS, will compile a final list that considers gender, age and country distribution, ensuring a diverse range of perspectives. 
This list will then be presented to and discussed with the Scientific Organising Committee.
In the initial stages of the process, suggestions for talks and their speakers were sometimes put to a vote by the Scientific Organising Committee.
While this voting process primarily considered the scientific and rhetorical aspects, it often failed to achieve a balanced distribution.

Being invited to give a talk, especially a review, at a major conference can be a valuable addition to a young scientist's CV. 
While an invited review is a significant undertaking, particularly for early-career researchers, it is crucial to present a balanced view of the subject matter and avoid overemphasising one's own findings. 
However, in order to keep the conference attractive to potential participants, it is essential that the majority of invited speakers are of the highest calibre and reputation, and that many are well known.

Regarding the X-Ray Universe conferences and certain workshops, the number of proposed oral presentations exceeds the available capacity. 
The Scientific Organising Committee selects the talks, reviewing each abstract submitted by three of its members.
The assignment of reviewers considers  potential conflict of interest, particularly if a committee member is an author or co-author of the abstract in question.  Consequently, abstracts within each discipline are usually reviewed by different individuals.
To ensure consistency in the evaluation process, the ratings provided by each reviewer are normalised to guarantee that the mean and standard deviation of the ratings remain consistent across all reviewers.
The mean of the normalised ratings is then calculated. 
The number of sessions is distributed in a way that ensures each topic receives the same oversubscription.
When scheduling the sessions, the talks with the highest mean normalised rating are typically selected for the first 70\% of slots.
For the remaining 30\%, a variety of additional factors were considered alongside the rating.

The proceedings of all workshops were published as refereed articles in the journal Astronomical Notes (e.g. 
\cite{2006AN....327..941S}, \cite{2025AN....34640093S}).
Consequently, only papers presenting original research were deemed suitable for inclusion in the proceedings.
In the case of review talks, the original contribution lies in synthesising and interpreting the topics under discussion.
This approach enabled the work of numerous early-career researchers to be published in a peer-reviewed journal for the first time.

\section{Data Rights}\label{dr}

The XMM-Newton proprietary periods have been designed to give PIs a fair and reasonable opportunity to explore their original ideas, as recommended by the OTAC, and to publish their XMM-Newton data in a peer-reviewed journal.

For individual observations, there is a one-year proprietary period, during which the data from an observation are made available only to the PI of the relevant proposal.
After this period, the data will be made available to the community.
The proprietary period begins when the data are made available to the PI in a usable form, i.e. when appropriate calibration and data processing are available. 
In the vast majority of cases, there is a delay of no more than one month between observation and data delivery. However, longer delays may occur in certain instances, such as when reprocessing is required due to a command failure during instrument setup or loss of downlink telemetry.

This is modified for proposals involving more than one observation of a single target, e.g. repeated observations for variability studies or observations that have been partially repeated due to high background radiation.
For all data delivered within an observation cycle for which the proposal  has been accepted by OTAC, the one-year retention period begins when the final data are made available to the PI.
Data taken after the end of the observation cycle for which the proposal has been accepted by OTAC will be treated as single observations; each dataset will receive a one-year proprietary period.

Data resulting from Multi-Year Heritage Programmes will not be assigned proprietary rights by default; however, the PS may grant these programmes data rights upon request.

The PS will decide on the data rights of unanticipated TOOs.
To ensure the proper analysis of XMM-Newton
data, the PS will normally grant a proprietary period of 0.5 years.

All data taken during a spacecraft slew will be immediately available to the community.

\section{Conclusions}\label{con}

 XMM-Newton's operational lifespan has already exceeded initial expectations,   with current fuel consumption rates allowing operations to continue well beyond 2032.
This prolonged scientific appeal is primarily due to the fundamental design of its payload, which incorporates six simultaneous observation instruments, as well as its robust mechanical and electrical structure. 
The arguments presented herein demonstrate that the discovery potential of a space mission such as XMM-Newton is subject to some inherent limitations, although this does not preclude significant scientific discoveries. 
The analysis further indicates that significant discoveries tend to be made within the first seven years of operation, after which point further progress requires prohibitively long exposure times. This pattern resembles that of large ground-based experiments in fundamental physics and may reflect the natural cycles of academic careers and shifting research priorities.
However, a key distinction is that space observatories have little or no capacity for replacement or upgrades, unlike ground-based facilities.
Nevertheless, the XMM-Newton observatory demonstrates that a mission can last more than 25 years and remain at the forefront of research in a specific area.
It is clear that scientific understanding undergoes substantial transformation over such extended periods. For instance, 
 research areas that were central at the project’s inception, such as high-resolution spectroscopy of cool stars, have declined in prominence. 
Conversely, emerging research domains such as the impact of X-rays on exoplanets and tidal disruption events were not even foreseen at launch. This highlights the difficulty of predicting scientific advances over a decade-long
timescale.

 The scientific rationale behind the expansion of the discovery space involved identifying opportunities 
 beyond the original scope of the mission.
These 
included facilitating large and very large observing programmes, conducting joint observations with new facilities and identifying novel transient phenomena that merited further investigation. 
Such changes enabled new observing strategies 
aligned with evolving scientific understanding, 
reflecting shifts in the research community and in broader social and behavioral patterns over a human lifetime.
A fundamental tenet of the XMM-Newton mission has been its openness to change, supported by committed engagement with the community through various means. These include studying and understanding work processes, actively monitoring scientific success indicators and soliciting feedback. This openness has involved the proactive adoption of guidelines and procedures, enabling the timely management of novel requests. Attention was also given to selecting the annual observing programme 
 through careful modifications to the review process, supporting research by small teams and early-career scientists, and   assisting observers.
Data analysis was facilitated through updated calibration and software.
Moreover, the professionalism and long-term commitment of the SOC personnel have been instrumental in enabling the mission to evolve in step with emerging scientific challenges. 
  Beyond the discoveries already made, more than 25 years of XMM-Newton operations ensure that the archive will continue to facilitate unexpected future findings, including long-term, time-domain studies. 

In the context of prospective observatory-type missions, the analysis presented here demonstrates the importance of incorporating a long-term perspective into the design of the spacecraft and its instruments from the outset. 
This approach should encompass fuel and maintenance requirements, ground hardware, and software components.
Although scientific rationale and observational scenarios are pivotal in the design process,
 scientific understanding and research inevitably evolve over time. 
Once launched, missions must be operated in a highly adaptable manner, demonstrating resilience to evolving scientific and social environments.

\hspace{1cm}

\textit{Acknowledgments}

The authors consider it a privilege to have contributed to XMM-Newton from shortly before its launch, and to have witnessed its scientific impact steadily grow over the past 25 years.   They are grateful to all SOC and MOC staff, the SSC and GOF teams, and everyone who helped bringing the mission to fruition, as well as to the XMM-Newton Users’ Group, the Time Allocation Committee, and all scientists who have proposed, analysed, and published XMM-Newton data.

The contributions of Pedro Rodríguez, who provided the histograms in Figures \ref{EvoNum} and \ref{EvoTime}, and Lucia Ballo, who assisted with internal databases, are gratefully acknowledged. The authors are also indebted to Michael Smith for insightful discussions on calibration constraints and helpful comments, and to Professor Phil Charles for his enthusiasm and invaluable assistance in refining the text. Constructive feedback from the anonymous referee significantly improved the comprehensiveness of the paper.

Using DeepL Pro and ChatGPT (OpenAI) facilitated the editing of the text and improvement of clarity.

\subsection{Bibliography}

\bibliography{main}

\end{document}